\documentclass[final,5p,times,twocolumn, sort&compress]{elsarticle}

\usepackage{lineno,hyperref}
\modulolinenumbers[5]

\usepackage{amsmath}
\usepackage{amsfonts}
\usepackage{amssymb,float}
\usepackage{mathtools}
\usepackage[utf8]{inputenc}
\usepackage{graphicx}
\usepackage{wasysym}
\usepackage{tabularx}
\usepackage{accents}
\usepackage{graphicx}
\usepackage{color, colortbl}
\usepackage[dvipsnames]{xcolor}
\usepackage{hyperref}
\usepackage{ulem}

\definecolor{Gray}{gray}{0.9}
\definecolor{LightCyan}{rgb}{0.88,1,1}

\hypersetup{
    colorlinks=true,
    linkcolor=blue,
    filecolor=magenta,      
    citecolor=red
}

\usepackage{subfigure} 
\usepackage{hyperref} 
\usepackage{multirow}

\newcommand{\be}{\begin{equation}}
\newcommand{\ee}{\end{equation}}
\newcommand{\bea}{\begin{eqnarray}}
\newcommand{\eea}{\end{eqnarray}}

\newcommand{\mpl}{M^2_{\rm  pl}}

\journal{}

\begin{document}

\begin{frontmatter}

\title{Looking out for the Galileon in the nanohertz gravitational wave sky}


\author[a]{Reginald Christian Bernardo\corref{mycorrespondingauthor}}
\cortext[mycorrespondingauthor]{Corresponding author}
\ead{rbernardo@gate.sinica.edu.tw}

\author[a,b]{Kin-Wang Ng}
\ead{nkw@phys.sinica.edu.tw}
\address[a]{Institute of Physics, Academia Sinica, Taipei 11529, Taiwan}
\address[b]{Institute of Astronomy and Astrophysics, Academia Sinica, Taipei 11529, Taiwan}

\begin{abstract}
We study the polarizations induced by the Galileon as a stochastic gravitational wave background in the cross correlated power in a pulsar timing array. Working within Galileon gravity, we first show that the scalar gravitational wave signature of the Galileon is encoded solely in its effective mass, which is controlled by the bare mass, conformal coupling, and a tadpole. Then, we study the phenomenology of the Galileon induced scalar polarizations and place observational constraints on these using the present NANOGrav data set. Our results feature longitudinal spatial correlation, indicative of a $10^{-22}$ eV {subluminal} Galileon, and show the Galileon polarizations as a statistically compelling source of the observed spatial correlation across millisecond pulsars, if there is any.
\end{abstract}


\end{frontmatter}




\noindent \textbf{\textit{Introduction.}} The North American Nanohertz Observatory for Gravitational Waves (NANOGrav) \cite{NANOGrav:2020bcs}{, Parkes Pulsar Timing Array \cite{Goncharov:2021oub}, and the European Pulsar Timing Array \cite{Chen:2021rqp}} observed a {strong evidence in favor a common spectrum process, but without any significant monopolar, dipolar, and quadrupolar spatial} correlation{s} between the power distributed over millisecond pulsars at different lines of sight at a reference frequency of {about} one per year. This outstanding result teases a departure from what should otherwise be {the quadrupolar} Hellings-Downs (HD) signal if the stochastic gravitational wave background (SGWB) is {dominated by} the transverse tensor modes, expected in general relativity \cite{Hellings:1983fr}. Such pulsar timing array (PTA) efforts complement the existing \cite{LIGOScientific:2021djp} and future \cite{LISA:2022kgy} gravitational wave observatories which operate at much higher frequencies, all the while providing an independent avenue to test our understanding of gravity in the nanohertz observational window \cite{Burke-Spolaor:2018bvk, Lentati:2015qwp, Shannon:2015ect, Janssen:2014dka, Romano:2019yrj}.

In this letter, we present observational constraints on a well featured scalar field theory -- the `covariant Galileon' \eqref{eq:gals} -- in the pulsar timing setting. In contrast with previous searches of non-Einsteinian polarizations highlighting transverse mode effects \cite{NANOGrav:2021ini, Chen:2021wdo, Chen:2021ncc}, we find a longitudinal correlation by letting the Galileon propagate freely away from the light cone. We utilize a formalism developed for PTA and astrometry observables \cite{Dai:2012bc, Qin:2018yhy, Qin:2020hfy}, and make the data constrain the parameter space of the Galileon. Our results are summarized in Table \ref{tab:margstats}, which speaks of a $10^{-22}$ eV {subluminal} Galileon{, sourcing the observed spatial correlation}.

\medskip

\noindent \textbf{\textit{Galileon gravity.}} The Galileon action can be written as \cite{Deffayet:2009wt, DeFelice:2010pv}
\be
\label{eq:gals}
\begin{split}
S[g, \phi] = & \int d^4 x \sqrt{-g} \bigg( \dfrac{\mpl}{2} R \left( 1 + \alpha \dfrac{\phi}{M_\text{pl}} \right) - \Lambda \\
& \phantom{gggg} - \dfrac{ \left( \partial \phi \right)^2 }{2} - \lambda^3 \phi - \dfrac{\mu^2 \phi^2}{2} - \dfrac{ \left( \partial \phi \right)^2 }{2 \kappa^3} \Box \phi \bigg)
\end{split}
\ee
where $g_{ab}$ is the metric, $R$ is the Ricci scalar, $\mpl = c^4/(8 \pi G)$ is the reduced Planck mass squared, $\phi$ is the scalar field, or the Galileon, $\left( \partial \phi \right)^2 = g^{ab} \left( \partial_a \phi \right) \left( \partial_b \phi \right)$, $\lambda$, $\mu$, and $\kappa$ are Galileon mass scales, $\alpha$ is a dimensionless conformal coupling constant, and $\Lambda$ is a cosmological constant. The various constants controlling the phenomenology of the Galileon are as follows: $\mu$ describes the bare mass of the scalar field, endowing the model with a Chameleon screening mechanism at high density environments \cite{deRham:2012az, Joyce:2014kja}; $\kappa$ is a braiding mass, equipping the model with a Vainshtein screening mechanism to hide the scalar field at small distances \cite{Brax:2011sv, Ali:2012cv, Andrews:2013qva}; $\lambda$ is a tadpole mass, which is able to accommodate a low energy vacuum state regardless of a Planck scale vacuum energy \cite{Appleby:2018yci, Appleby:2022bxp}; and $\alpha$ is a dimensionless constant measuring the conformal coupling between the tensor and scalar modes \cite{Brax:2014vla}. The theory \eqref{eq:gals} is `shift symmetric' in the limit $\mu = 0$ in the sense $\phi \rightarrow \phi + C$ for arbitrary constant $C$ does not alter the dynamics. Shift symmetry protects the massless Galileon from radiative corrections \cite{Burrage:2010cu}. In the presence of a nonzero bare mass $\mu$, a shift transformation instead adds to the mass of the tadpole which sources the scalar field dynamics. The tensor modes in \eqref{eq:gals} propagate at the speed of light in agreement with gravitational wave astronomy constraints \cite{LIGOScientific:2017vwq, LIGOScientific:2021sio}.

The theory is satisfied by a static vacuum solution which includes Minkowski spacetime and Schwarzschild black hole. We proceed to study the gravitational perturbations on this background.

\medskip

\noindent \textbf{\textit{Perturbations.}} We reveal the propagating degrees of freedom of the Galileon by looking at its perturbations.

We fix the residual gauge degrees of freedom by working in the synchronous gauge:
\be
\begin{split}
ds^2 = -dt^2 + ( \delta_{AB} & - 2 \psi \delta_{AB} + 2 D_A D_B E \\
& + 2 D_{(A} E_{B)} + 2 E_{AB} ) dx^A dx^B \,,
\end{split}
\ee
where $\psi$ and $E$ are scalars, $E_A$ is a transverse vector ($D_A E^A = 0$), and $E_{AB}$ is a transverse-traceless (TT) tensor ($D^A E_{AB} = 0$ and $\delta^{AB} E_{AB} = 0$). Above, $D$ is the covariant derivative operator defined by the spatial metric. We work with the scalar field
\be
\phi = \varphi + \delta \phi \,.
\ee
A simple counting of components, $\psi (+1), E (+1), E_B (+2), E_{AB} (+2)$, shows that the gauge is now fully fixed. We turn off the vector modes ($E_B = 0$) as they are nondynamical and diluted by the expansion in the context of Galileon gravity. 

Using the constraint equations, it can be shown that the linearized scalar field equation simplifies to a massive Klein-Gordon equation,
\be
D^2 \psi - \ddot{\psi} - m_\text{eff}^2 \psi = 0 \,,
\ee
where the effective mass $m_\text{eff}$ is given by
\be
\label{eq:gal_mass}
m_\text{eff}^2 = \mu^2 \dfrac{\left( 1 - \alpha \lambda^3 / (M_\text{pl} \mu^2) \right) }{ \left( 1 + (3\alpha^2/2) - \alpha \lambda^3/(M_\text{pl} \mu^2) \right)} \, .
\ee
This shows that the scalar perturbations are controlled by three parameters: $\mu$, $\alpha$ and $\lambda^3/ \left( M_\text{pl} \mu^2 \right)$. The conformal coupling and the tadpole changes the mass of the scalar modes relative to its bare mass. We may traditionally avoid tachyonic instability as long as the effective mass-squared is positive. This narrows down the parameter space of the conformal coupling and the tadpole. We safely consider the stable region for this work.

The propagating scalar modes of the Galileon therefore satisfy a massive dispersion relation
\be
\label{eq:dispersion_scalar}
\omega^2 = k^2 + m_\text{eff}^2
\ee
where $\omega = 2\pi f$ and $\vec{k} = k \hat{k}$ are the frequency and wave number, respectively. Such massive dispersion relation leads to a group velocity $v = d\omega/dk = k/\omega$ that is reciprocal of the phase velocity $v_\text{ph} = 1/v = \omega/k$. Combining the above results, we find that the scalar perturbations in the metric bring in scalar transverse and longitudinal polarizations, which we put out by writing the metric perturbation as
\be
\label{eq:metric_scalar_gal}
h_{AB} \propto \left( \varepsilon_{AB}^{\text{ST}} + \dfrac{1-v^2}{\sqrt{2}} \varepsilon_{AB}^{\text{SL}} \right) \tilde{\psi}\left(k\right) e^{i(\vec{k} \cdot \vec{x} - \omega t)} \,,
\ee
where the polarization tensors $\varepsilon_{AB}$ are normalized such that $\varepsilon^{P, AB} \varepsilon^{P'}_{AB} = 2 \delta^{PP'}$ where $P = \text{ST}, \text{SL}$ stands for `scalar transverse' and `scalar longitudinal' polarizations. This lets us identify the ratio of the SL to ST amplitudes as $\left(1 - v^2\right)/\sqrt{2} = m_\text{eff}^2/\left(\sqrt{2}\omega^2\right)$. The longitudinal response is sourced by a nonzero scalar effective mass.

It is straightforward to bring the tensor modes $E_{AB}$ into the picture. This does not enter the Hamiltonian nor the momentum constraints. Rather, its influence can be found in the spatial component of the metric equation which turns out to be
\be
\ddot{E}_{AB} - D^2 E_{AB} = 0 \,,
\ee
which shows that tensor modes propagate on the light cone. This gives rise to the TT polarizations which were observed by the present ground-based gravitational wave observatories \cite{LIGOScientific:2021djp}. In the context of a SGWB, the tensor polarizations source the HD correlation in a PTA \cite{Hellings:1983fr, Jenet:2014bea}.

{To end the section, we add that while Galileon gravity contains scalar and tensor modes, the mechanisms for exciting these vary depending on the astrophysical configuration of its sources, e.g., tensor modes are suppressed in a spherical collapse, but are appreciably produced by a binary system. With this in mind, it is acceptable to consider the scalar and tensor modes individually in the correlations, also noting that the tensor excitations propagate at the speed of light, indistinguishable from their general relativity counterpart. On the other hand, that theory brings in a massive scalar that can be distinctly associated with the Galileon and produces clear cut correlations in the SGWB. We proceed to look for this observational signature.}

\medskip

\noindent \textbf{\textit{Scalar polarizations.}} We present the relevant results emerging from the total angular momentum formalism \cite{Dai:2012bc, Qin:2018yhy, Qin:2020hfy}, focusing on the scalar polarizations and pulsar timing, and considering finite pulsar distances.

Provided an isotropic background, we expand the metric perturbation as
\be
\label{eq:tam_series}
h_{AB} = \int dk \dfrac{k^2}{\left(2\pi\right)^3} 4\pi i^l h_{lm}^\sigma(k) \Psi^{\sigma, k}_{(lm)AB}(\vec{x}) e^{-i \omega_\sigma (k) t}
\ee
where $\Psi^{\sigma, k}_{(lm)AB}$ are total angular momentum waves of polarization $\sigma$ \cite{Dai:2012bc}. With this, we compute the power spectra of the correlations from a SGWB, leading to
\be
\label{eq:power_spectra}
C^\sigma_l \propto 32 \pi^2 F^\sigma_l \left( F^{\sigma}_l \right)^* \,,
\ee
{where} $F^\sigma_l$ are the projection factors for a polarization $\sigma$ {for the PTA observable}. This is tabulated for the group velocity dependent scalar, vector, and tensor modes in \cite{Qin:2020hfy}.

A modification we invoke is to keep the pulsars at realistic finite distances. This enters the formalism as we put in \eqref{eq:tam_series} into the time integral of the pulsar timing residual. After a straightforward calculation, this leads to nearly the same projection factors as \cite{Qin:2020hfy} except that the upper limit of the integration is kept at a finite distance. The generalized projection factors turn out to be
\begin{equation}
    F_l(fD) = - \dfrac{i}{2} \int_0^{2\pi fDv} \dfrac{dx}{v} \ e^{ix/v} R_l\left(x\right) \,,
\end{equation}
where the $R_l(x)${'}s, involving the spherical Bessel functions $j_l(x)$, are given in \cite{Qin:2020hfy}, that is, $R_l^\text{SL}(x) = j_l''(x)$ for the SL polarization and $R_l^\text{ST}(x) = -\left( R_l^\text{SL}(x) + j_l(x) \right)/\sqrt{2}$ for the ST polarization. {A notable distinction due to keeping finite distances are the boundary terms, $F_l\left(fD\right) \propto e^{2\pi i fD} \left( j_l'\left(2\pi fD v\right) - \dfrac{i}{v} j_l\left(2\pi fD v\right) \right)$, which otherwise vanish for $l \geq 2$ in the infinite distance limit \cite{Qin:2020hfy}.}

We find that the finite distance modification influences the small angular scales, but since most of the observed pulsars are separated by quite a few degrees in the sky, the overall modification turns out to be mild \cite{Ng:2021waj, Chu:2021krj}. {Nonetheless, keeping finite pulsar distances contain about half the power in small scales, providing richer physics, which would otherwise be lost when taking the infinite distance limit.} Another practical consequence of keeping the pulsars at finite distance is that all of the modes become defined for arbitrary group velocities.

We tease the influence of realistic finite pulsar distances on the power spectra and the overlap reduction function (ORF) which measures the average correlated power between a pair of pulsars in the sky. The ORF for an isotropic SGWB with a power spectra $C_l$ can be shown to be
\begin{equation}
\label{eq:orf_gen}
    \Gamma_{ab} \left( { \zeta, fD_i } \right) = \sum_l \dfrac{2l+1}{4\pi} C_l P_l\left( \cos \zeta \right) \,,
\end{equation}
where the pulsars are located at a distance $D_i$, {$\zeta$ is the angular separation between a pair of pulsars}, and $P_l(x)$ are the Legendre polynomials.

We remind that in theories with a propagating scalar degree of freedom, as in the Galileon, the amplitudes of the scalar modes would often be constrained with respect to each other by the effective mass. Conversely, a nonzero effective mass would imply the presence of both modes in a mixture in a SGWB. We constrain these Galileon polarizations using PTA observation.

\medskip

\noindent \textbf{\textit{Galileon in the nHz GW sky.}} We write down the metric perturbation
\be
\label{eq:metric_scalar_gal_tam}
\begin{split}
h_{AB} = \int \dfrac{dk \, k^2}{ 2\pi^2}
& i^l Y_{lm}(\hat{k}) \bigg( \Psi^{\text{ST}, k}_{(lm)AB}(\vec{x}) + \dfrac{1 - v^2}{\sqrt{2}} \Psi^{\text{SL}, k}_{(lm)AB}(\vec{x}) \bigg) e^{-i \omega (k) t} \,,
\end{split}
\ee
where $\omega(k)$ is given by the massive scalar dispersion relation \eqref{eq:dispersion_scalar}. Our goal is to constrain each scalar polarization and their combination as the Galileon \eqref{eq:metric_scalar_gal_tam} using PTA data \cite{NANOGrav:2020bcs}.

We treat the pulsars at an average distance $fD = 100$, corresponding to an astronomical distance of about thirty parsecs. This practical assumption makes all of the multipoles defined, and influences small angular correlations irrelevant for the present data set \cite{NANOGrav:2021ini}. This way, we are left with two free parameters -- the velocity $v$ and the {GW} amplitude-squared $A^2$ -- to constrain with observation. We calculate the ORF, {$\Gamma_{ab}$,} provided the velocity $v$, and then compare the model $A^2 \times {\Gamma_{ab}}$ with the average angular distribution of cross correlated power in PTA measured by NANOGrav \cite{NANOGrav:2020bcs}. We consider three models: ST for scalar transverse, SL for scalar longitudinal, and $\phi$ for the Galileon. In calculating the ORF, we take the first thirty multipoles, resolving ${\zeta} = 180^\circ / l_\text{max} = 6^\circ$, which is sufficient for the observed pulsars separated by about ten degrees at the least{, and normalize each curve with respect to the {TT} one} {given by $\Gamma_{ab}(\zeta) = \delta_{ab}/2 + \Gamma_{ab}^{\text{HD}}(\zeta)$ where $\Gamma_{ab}^\text{HD} = 1/2$ at $\zeta = 0^\circ$}. For comparison, we also consider the {GW}{-like} monopole {($\Gamma_{ab} = \delta_{ab}/2 + 1/2$) \cite{NANOGrav:2021ini}}{,} and the HD correlation, both of which are described by a single free parameter, $A^2$. {We additionally consider the characteristic GW amplitude squared of the common spectrum process, $A_\text{CP}^2 = \left( 3.68 \pm 1.58 \right) \times 10^{-30}$ \cite{NANOGrav:2020bcs}, as an autocorrelation data point, $A_\text{CP}^2 = A^2 \Gamma_{aa}$ \cite{NANOGrav:2021ini}. In doing the $\Gamma_{aa}$ computation, we have included a thousand multipoles, guaranteeing the convergence of the sum \eqref{eq:orf_gen}, and have confirmed the result with the real space formalism \cite{NANOGrav:2021ini}}.

We mention that the current signal detection routines utilized in PTA analysis were optimized for the HD curve \cite{NANOGrav:2020bcs}, among the other correlations considered so far \cite{NANOGrav:2021ini}, which has no free parameters other than the overall characteristic GW strain amplitude. However, the Galileon correlations we are interested in are described by the mass scales of the theory that we intend to constrain using the PTA data. We therefore rely on standard Bayesian analysis implemented in cosmological parameter estimation \cite{Trotta:2008qt}, utilizing community codes \cite{Lewis:2019xzd, Torrado:2020dgo} that are widely employed in constraining cosmological models.

We perform a Markov chain Monte Carlo analysis over each model, with flat priors $v \in [0.01, 0.99]$ and $A^2 \in [0.1, {3}0] \times 10^{-30}$, and assess the statistical performance of each compared to the {GW} monopole. We consider the {likelihood function $\mathcal{L}$} given by
\begin{equation}
    {\log \mathcal{L}} \propto {-\dfrac{1}{2}} \sum_{{\text{pulsar pairs}}} \left( \dfrac{ \left( A^2 \Gamma_{ab} \right)_\text{NG} - { \left(A^2 \Gamma_{ab} \right)_\text{model} } }{\Delta \left( A^2 \Gamma_{ab} \right)_\text{NG}} \right)^{{2}} \,,
\end{equation}
where $\left( A^2 \Gamma_{ab} \right)_\text{NG} \pm \Delta \left( A^2 \Gamma_{ab} \right)_\text{NG}$ are the average cross correlated power across the PTA{.} {Then, we consider the chi-squared $\chi^2 \sim - 2 \log \hat{ \mathcal{L} }$ as a bare measure of statistical performance, where $\hat{\mathcal{L}}$ is the best fit likelihood.} We additionally consider the Akaike information criterion ($\text{AIC} = 2p - 2 \log \hat{\mathcal{L}}$) and Bayesian information criterion ($\text{BIC} = p \ln n - 2 \log \hat{\mathcal{L}}$) where $p$ is the number of parameters estimated, and $n$ is the size of the data set {\cite{Trotta:2008qt, Liddle:2007fy}}. Both AIC and BIC penalize models with more free parameters {and so are good lookouts for overfitting}. To be concrete, $p = 2$ for the scalar polarizations and the Galileon, $p = 1$ for the {GW} monopole and HD correlation, and $n = 15$ for the observation size.

\begin{figure}[h!]
\center
	\includegraphics[width = 0.48 \textwidth]{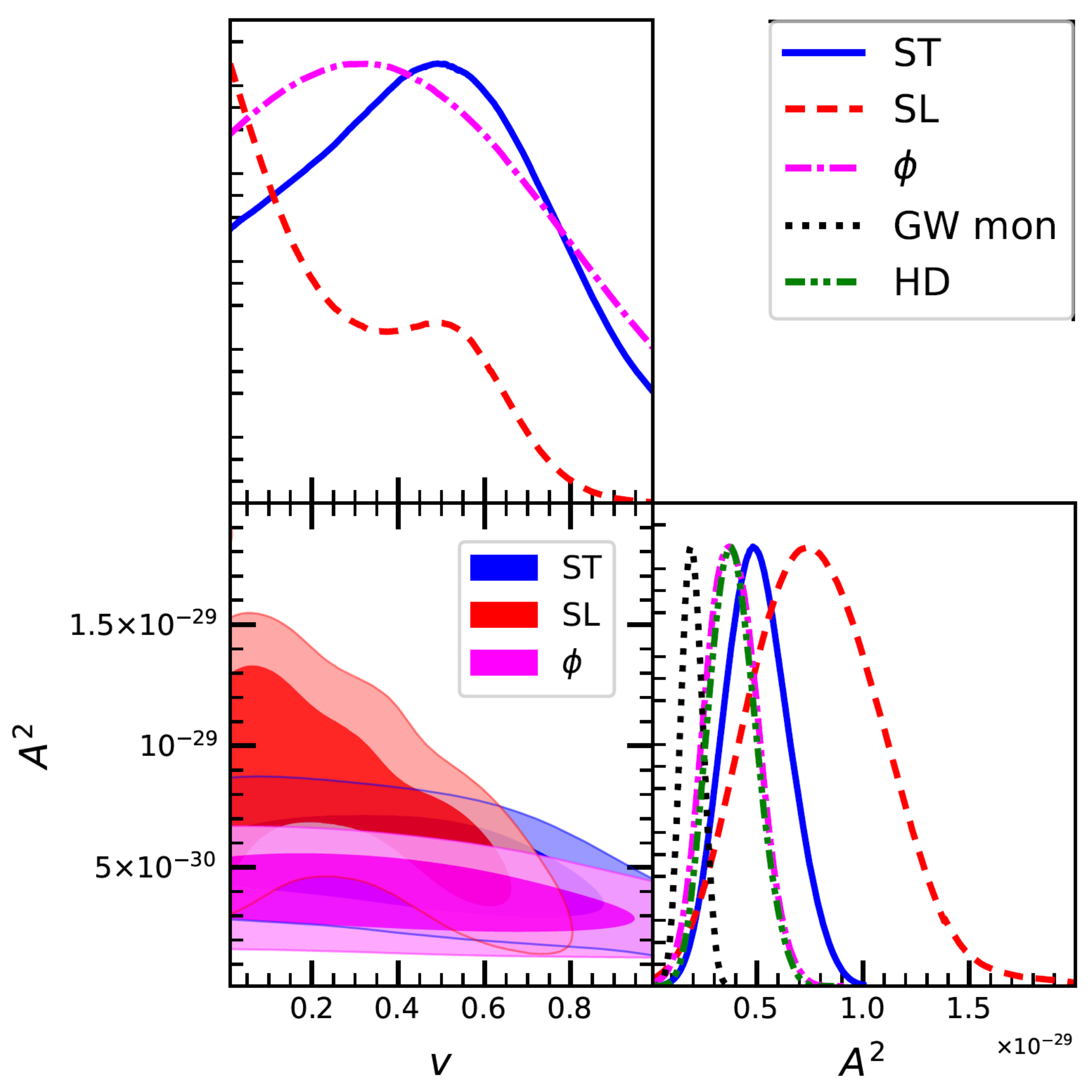}
\caption{Posteriors of the velocity ($v$) and the amplitude-squared ($A^2$) for each polarization (ST and SL) and the Galileon ($\phi$) constrained by PTA observation \cite{NANOGrav:2020bcs}. The black-dotted line results from the {GW} monopole{, while the green dot-dot-{dashed} line comes from the HD correlation}.}
\label{fig:posts}
\end{figure}

{We sample the parameter space of each model, each time starting at a different point, and take a Gelman-Rubin convergence criterion $R - 1 = 10^{-2}$ to decide the end of the sampling.} The posteriors for the velocity and amplitude are shown in Figure \ref{fig:posts} for the scalar polarizations, the Galileon, {the HD curve,} and the {GW} monopole.

For each model, we find that the amplitudes are constrained; however, the velocity {is generally not. Nonetheless, a recurring feature of the Galileon polarizations is that subluminal velocities are preferred, should such degrees of freedom be pronounced in the nanohertz GW sky.} The marginalized statistics for each model, and the assessment of their significance are presented in Table \ref{tab:margstats}.

\begin{table}[h!]
    \centering
    \caption{Marginalized statistics for the ST, SL, and the Galileon ($\phi$) constrained by PTA observation \cite{NANOGrav:2020bcs}. Results for the HD correlation and the {GW} monopole ({GW} mon.) are presented for comparison. The performance statistics (chi-squared, AIC, and BIC \cite{Trotta:2008qt, Liddle:2007fy}) are relative to the {GW} monopole, or that a {negative} value means statistical preference over the {GW} monopole.}
    \resizebox{0.48\textwidth}{!}{%
    \begin{tabular}{|c|c|c|c|c|c|} \hline
    mode & \phantom{11111} $v$ \phantom{11111} & $A^2$ [$\times 10^{-30}$] & $\Delta \chi^2$ & $\Delta$AIC & $\Delta$BIC \\ \hline \hline 
    ST & \phantom{$\dfrac{1}{1}$} $0.46 \pm 0.24$ \phantom{$\dfrac{1}{1}$} & $5.0^{+1.4}_{-1.6}$ & $-1.47$ & $0.53$ & $1.24$ \\ \hline
    SL & \phantom{$\dfrac{1}{1}$} $< 0.44$ \phantom{$\dfrac{1}{1}$} & $7.9^{+2.8}_{-3.4}$ & $-3.92$ & $-1.92$ & $-1.21$ \\ \hline
    \rowcolor{Gray}
    ${\boldsymbol \phi}$ & \phantom{$\dfrac{1}{1}$} $\mathbf{0.44^{+0.15}_{-0.42}}$ \phantom{$\dfrac{1}{1}$} & $\mathbf{3.8 \pm 1.2}$ & $\mathbf{-2.91}$ & $\mathbf{-0.91}$ & $\mathbf{-0.20}$ \\ \hline 
    HD & \phantom{$\dfrac{1}{1}$} $v = 1$ \phantom{$\dfrac{1}{1}$} & $3.9 \pm 1.1$ & $1.66$ & $1.66$ & $1.66$ \\ \hline 
    {GW} mon. & \phantom{$\dfrac{1}{1}$} $---$ \phantom{$\dfrac{1}{1}$} & $1.94 \pm 0.48$ & $0$ & $0$ & $0$ \\ \hline 
    \end{tabular}}
    \label{tab:margstats}
\end{table}

{
In assessing the goodness of fit of the various GW correlation models, we consider the GW-like monopole as a baseline, which is reasonable given its significance in the present data, and so opted to present these statistical measures as {$\Delta {\rm XIC} = {\rm XIC}_{\rm model} - {\rm XIC}_{\rm GW mon}$}, where XIC is either $\chi^2$ or AIC or BIC. This makes the interpretation transparent, e.g., if a model is favored over the GW monopole, ${\rm XIC}_{\rm model} < {\rm XIC}_{\rm GW mon}$ or {$\Delta {\rm XIC} < 0$}. In short, ({$-$}) implies favorability and ({$+$}) otherwise in Table \ref{tab:margstats}.
}

\begin{figure}[h!]
\center
	\includegraphics[width = 0.48 \textwidth]{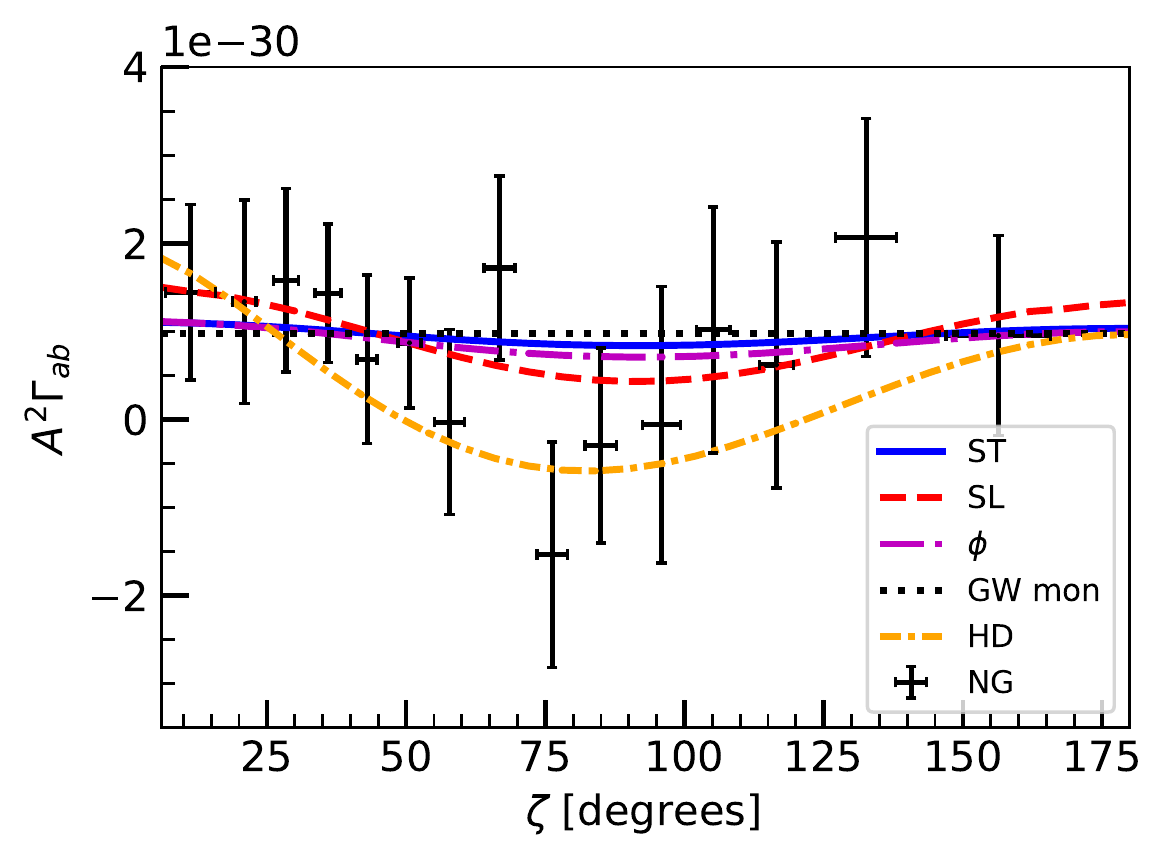}
\caption{Mean curves of each polarization (ST and SL) and the Galileon ($\phi$) constrained by PTA observation (NG, for NANOGrav). The best fit {GW} monopole (mon) and HD correlation are also presented for visual comparison.}
\label{fig:mean_curves}
\end{figure}

First off, we echo the earlier result that the HD correlation is weaker compared to the {GW} monopole \cite{NANOGrav:2020bcs}. This time, it is reflected in a Bayesian analysis through the chi-squared, AIC, and BIC, as shown in Table \ref{tab:margstats}, which are {consistently} in the positive direction compared with the {GW} monopole. Both scalar polarizations and the Galileon also statistically perform better compared to the HD correlation. As far as the chi-squared is concerned, all of these {alternative SGWB scenarios} do better than the {GW} monopole in representing the data. However, when the AIC and the BIC are considered, taking into account the number of parameters, the ST polarization come{s} with {a} positive measure{, contrary to what was found for luminal modes \cite{NANOGrav:2021ini, Chen:2021wdo, Chen:2021ncc}}. Granted, these positive values are mild and not enough to disfavor their correlations compared with the {GW} monopole. On the other hand, even with the AIC and BIC measures, the SL polarization {and the Galileon} by {themselves are} able to compete, even surpass, {with} the fit provided by the {GW} monopole. Admittedly, this is mild or irrelevent given the uncertainty of the present data, but it hints at unorthodox possibilities -- in this case, the presence of the Galileon -- when it comes to nanohertz {GW} science. {T}he mean curves for each model are presented in Figure \ref{fig:mean_curves} together with the data set.

We find here (Figure \ref{fig:mean_curves}) the {GW} monopole, which was previously found to be the strongest signal in this data set \cite{NANOGrav:2021ini, Chen:2021wdo, Chen:2021ncc}, and the HD correlation. The ST and the Galileon polarizations draw close by the {GW} monopole. We recall both of these are {nearly as competitive as} the {GW} monopole. On the other hand, the SL polarization can be seen to form a smooth curve resembling what is between the {GW} monopole and the HD correlation. Echoing the statistical measures in Table \ref{tab:margstats}, this best fit curve of the SL polarization can be seen to be slightly statistically preferred over the {GW} monopole.

{
We clarify that the ST and SL modes, independently, do not make sense in a gravity framework. Another way of looking at this is that they only show up physically as a mixture (depending on the GW speed) as shown in \cite{Qin:2020hfy} for $f(R)$ gravity and in this work for the Galileon. Nonetheless, if any, the SL and ST correlations statistics give more impact to the HD/Galileon's significance by providing extra phenomenological models to compare the physical correlations with, in addition to paying homage to earlier important works, e.g., \cite{NANOGrav:2021ini, Chen:2021wdo}, that have considered these modes separately and paved the current path we are threading.
}

\begin{figure}[h!]
\center
	\includegraphics[width = 0.48 \textwidth]{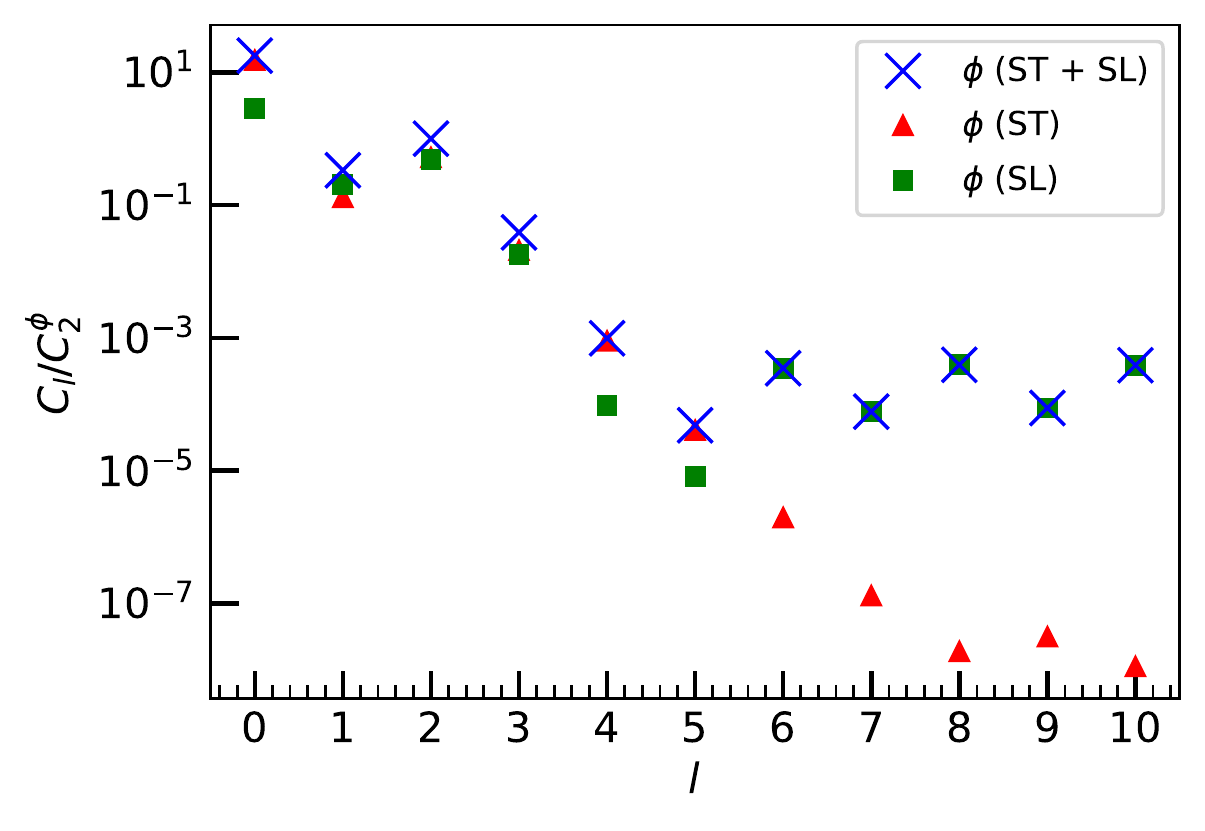}
\caption{{The first ten angular power spectrum multipoles of the best fit Galileon $\phi$ (Eq. \eqref{eq:metric_scalar_gal_tam}) together with its ST and SL components.}}
\label{fig:mean_cls}
\end{figure}

{
The best fit angular power spectrum multipoles of the Galileon as well as its transverse and longitudinal components are shown in Figure \ref{fig:mean_cls}.
}

{
This reveals the distinct subluminal scalar SGWB power spectrum profile (first recognized in \cite{Qin:2020hfy}) where in order the most significant contributions come from the monopole ($C_0$), quadrupole ($C_2$), dipole ($C_1$), octupole ($C_3$) which together shape the correlation at the largest angles, $\zeta \gtrsim 60^\circ$. It is also worth noting that this trend holds for either the transverse and longitudinal components of the Galileon. However, at lower angles, it can be seen that the longitudinal contributions now become more prominent than the transverse ones by several orders of magnitude, which can be traced to a finite distance effect \cite{Ng:2021waj}. This teases the potential impact of accounting for astrophysical pulsar distances in PTA correlations analysis, particularly for unmasking (or ruling out) the existence of longitudinal scalar GW polarizations.
}

{We emphasize that we refer to the GW monopole \cite{NANOGrav:2021ini} throughout this work, and \textit{not} the systematic monopole, which is disfavored by the NANOGrav and more so by the Parkes Pulsar Timing Array data \cite{NANOGrav:2020bcs, Goncharov:2021oub}. Indeed, the systematic monopole comes with a signal-to-noise ratio exceeding the HD and the dipole, but at face value, this should not be taken as evidence in favor of the monopole. Rather, it means conservatively that the spatially correlated monopole cannot be clearly separated from the spatially uncorrelated common process, pointing to the limitations in the optimal statistic analysis.}

The results of the data analysis show that the Galileon polarizations represent the current PTA data better than the quadrupolar HD correlation. Of course, the error bars on the current data set are too large to make stronger conclusions at the moment, but we may at least take the result to look forward for new science in the nanohertz {GW} regime. \\

\medskip

\noindent \textbf{\textit{Discussion.}} We emphasize in constrast with previous analyses \cite{NANOGrav:2021ini, Chen:2021wdo, Chen:2021ncc}, we let the scalar modes off the light cone, or rather that we constrained the velocity using the available data. This led to stronger longitudinal correlation compared with the transverse ones. For the SL mode off the light cone, we find that the {autocorrelation} {varies as} {$\Gamma_{aa} = (0.33, 0.71)$} for $0.2 \lesssim v \lesssim 0.6$, with only mild dependence on the pulsar distance, whereas from Figure \ref{fig:mean_curves}{,} {$\Gamma_{ab} \simeq 0.20$} {at} ${\zeta}=0^\circ$ and {$A^2 = 7.9 \times 10^{-30}$}. This is in stark contrast with the $v = 1$ case \cite{Chamberlin:2011ev}{,} which is disfavored by the current data \cite{NANOGrav:2021ini}.

\begin{figure}[h!]
    \centering
    \includegraphics[width = 0.48 \textwidth]{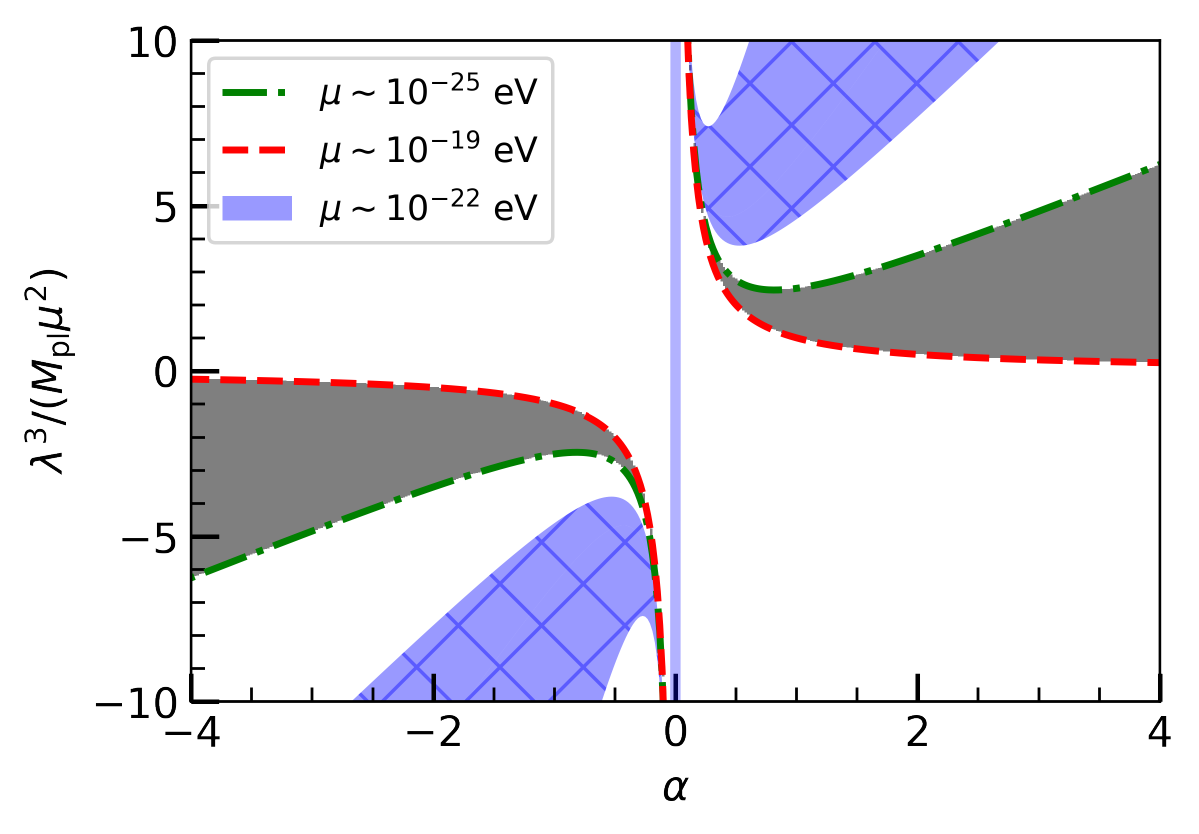}
    \caption{{Constraints on the parameter space of the Galileon \eqref{eq:gals}. The {gray} region is excluded by tachyonic instability. The blue hatched region, red dashed line, and green dash-dotted line present the pulsar timing array constraints based on different scales of the Galileon bare mass $\mu$.}}
    \label{fig:tachyon}
\end{figure}

We draw the discussion back to the Galileon and see the implications of the constraints discussed previously. In standard units, the group velocity can be written as
\begin{equation}
    v = c \left( 1 - \left( \dfrac{m_\text{eff} c^2}{hf} \right)^2 \right)^{1/2} \,.
\end{equation}
While the error bars are yet too large to draw a precise calculation, we may estimate the Galileon mass resulting from $v \sim c/2$, as supported by the present PTA data. This leads to $m_\text{eff} c^2 \sim \sqrt{3} hf / 2$. Using the NANOGrav reference frequency $f = 1 \, \text{yr}^{-1}$, we obtain $m_\text{eff} c^2 \sim 10^{-22}$ eV. This nonzero effective mass \eqref{eq:gal_mass} constrain{s} the Galileon's bare mass, conformal coupling, and tadpole weight, in addition to the theoretical constraint on eliminating potential tachyonic instability {(Figure \ref{fig:tachyon})}.

{The blue region in Figure \ref{fig:tachyon} shows the surviving parameter space, at sixty eight percent confidence, should the bare mass be of the order of the effective mass.} This may be taken for one to mean that the effective mass is just the bare mass, in which case the conformal coupling and the tadpole vanish. However, a heavier bare mass may be needed to pass Solar system constraints \cite{deRham:2012az, Joyce:2014kja}. The conformal coupling and the tadpole can act in this regard to still make the theory consistent with pulsar timing observations. {Such is shown by the red dashed and green dash{-}dotted curves in Figure \ref{fig:tachyon}, even though these may be dangerously close to the tachyonic region (in the {gray region}).} This may be particularly indicative of the tadpole since the conformal coupling can be strongly constrained by other observations. We add that this also makes the case of the Galileon, in comparison with $f(R)$ gravity \cite{Qin:2020hfy}, which may be immediately ruled out by such effective mass estimate after considering Solar system constraints. Further, we highlight that the braiding can also be expected to hide the Galileon in such environments where nonlinearities settle in \cite{Brax:2011sv, Ali:2012cv, Andrews:2013qva}.

We end on an optimistic note. The variety of scientific outcomes anchored on astronomical data is something to look forward to. Despite the size of the uncertainty so far, our result hints at such possibilities in nanohertz {GW} science, this time, teasing the presence of the Galileon. This warrants a detailed study of the parameter space of Galileon gravity, which we shall discuss elsewhere.

\medskip

\noindent {\textbf{\textit{Supplementary files.}} We present supplementary notes detailing the theoretical derivation of the Galileon polarizations, a brief analysis of its phenomenology, and a python notebook disclosing the modelling and data analysis done in this work in \href{https://github.com/reggiebernardo/galileon_pta}{GitHub} \cite{rbgithub}.}

\medskip

\noindent \textbf{\textit{Acknowledgments.}} The authors thank Che-Yu Chen for discussion over a preliminary draft, and Sarah Vigeland and Aaron Johnson for helping them work with the jupyter notebooks of NANOGrav. The authors acknowledge the use of community codes GetDist \cite{Lewis:2019xzd} and Cobaya \cite{Torrado:2020dgo} for data analysis. This work was supported in part by the Ministry of Science and Technology (MOST) of Taiwan, Republic of China, under Grant No. MOST 110-2112-M-001-036.



\begin{thebibliography}{10}
\expandafter\ifx\csname url\endcsname\relax
  \def\url#1{\texttt{#1}}\fi
\expandafter\ifx\csname urlprefix\endcsname\relax\def\urlprefix{URL }\fi
\expandafter\ifx\csname href\endcsname\relax
  \def\href#1#2{#2} \def\path#1{#1}\fi

\bibitem{NANOGrav:2020bcs}
Z.~Arzoumanian, et~al., {The NANOGrav 12.5 yr Data Set: Search for an Isotropic
  Stochastic Gravitational-wave Background}, Astrophys. J. Lett. 905~(2) (2020)
  L34.
\newblock \href {http://arxiv.org/abs/2009.04496} {\path{arXiv:2009.04496}},
  \href {http://dx.doi.org/10.3847/2041-8213/abd401}
  {\path{doi:10.3847/2041-8213/abd401}}.

\bibitem{Goncharov:2021oub}
B.~Goncharov, et~al., {On the Evidence for a Common-spectrum Process in the
  Search for the Nanohertz Gravitational-wave Background with the Parkes Pulsar
  Timing Array}, Astrophys. J. Lett. 917~(2) (2021) L19.
\newblock \href {http://arxiv.org/abs/2107.12112} {\path{arXiv:2107.12112}},
  \href {http://dx.doi.org/10.3847/2041-8213/ac17f4}
  {\path{doi:10.3847/2041-8213/ac17f4}}.

\bibitem{Chen:2021rqp}
S.~Chen, et~al., {Common-red-signal analysis with 24-yr high-precision timing
  of the European Pulsar Timing Array: inferences in the stochastic
  gravitational-wave background search}, Mon. Not. Roy. Astron. Soc. 508~(4)
  (2021) 4970--4993.
\newblock \href {http://arxiv.org/abs/2110.13184} {\path{arXiv:2110.13184}},
  \href {http://dx.doi.org/10.1093/mnras/stab2833}
  {\path{doi:10.1093/mnras/stab2833}}.

\bibitem{Hellings:1983fr}
{{Hellings}, R.~W. and {Downs}, G.~S.}, {Upper limits on the isotropic
  gravitational radiation background from pulsar timing analysis.}, Astrophys.
  J. Lett. 265 (1983) L39--L42.
\newblock \href {http://dx.doi.org/10.1086/183954} {\path{doi:10.1086/183954}}.

\bibitem{LIGOScientific:2021djp}
R.~Abbott, et~al., {GWTC-3: Compact Binary Coalescences Observed by LIGO and
  Virgo During the Second Part of the Third Observing Run}\href
  {http://arxiv.org/abs/2111.03606} {\path{arXiv:2111.03606}}.

\bibitem{LISA:2022kgy}
K.~G. Arun, et~al., {New Horizons for Fundamental Physics with LISA}\href
  {http://arxiv.org/abs/2205.01597} {\path{arXiv:2205.01597}}.

\bibitem{Burke-Spolaor:2018bvk}
S.~Burke-Spolaor, et~al., {The Astrophysics of Nanohertz Gravitational Waves},
  Astron. Astrophys. Rev. 27~(1) (2019) 5.
\newblock \href {http://arxiv.org/abs/1811.08826} {\path{arXiv:1811.08826}},
  \href {http://dx.doi.org/10.1007/s00159-019-0115-7}
  {\path{doi:10.1007/s00159-019-0115-7}}.

\bibitem{Lentati:2015qwp}
L.~Lentati, et~al., {European Pulsar Timing Array Limits On An Isotropic
  Stochastic Gravitational-Wave Background}, Mon. Not. Roy. Astron. Soc.
  453~(3) (2015) 2576--2598.
\newblock \href {http://arxiv.org/abs/1504.03692} {\path{arXiv:1504.03692}},
  \href {http://dx.doi.org/10.1093/mnras/stv1538}
  {\path{doi:10.1093/mnras/stv1538}}.

\bibitem{Shannon:2015ect}
R.~M. Shannon, et~al., {Gravitational waves from binary supermassive black
  holes missing in pulsar observations}, Science 349~(6255) (2015) 1522--1525.
\newblock \href {http://arxiv.org/abs/1509.07320} {\path{arXiv:1509.07320}},
  \href {http://dx.doi.org/10.1126/science.aab1910}
  {\path{doi:10.1126/science.aab1910}}.

\bibitem{Janssen:2014dka}
G.~Janssen, et~al., {Gravitational wave astronomy with the SKA}, PoS AASKA14
  (2015) 037.
\newblock \href {http://arxiv.org/abs/1501.00127} {\path{arXiv:1501.00127}},
  \href {http://dx.doi.org/10.22323/1.215.0037}
  {\path{doi:10.22323/1.215.0037}}.

\bibitem{Romano:2019yrj}
J.~D. Romano, {Searches for stochastic gravitational-wave backgrounds}, 2019.
\newblock \href {http://arxiv.org/abs/1909.00269} {\path{arXiv:1909.00269}}.

\bibitem{NANOGrav:2021ini}
Z.~Arzoumanian, et~al., {The NANOGrav 12.5-year Data Set: Search for
  Non-Einsteinian Polarization Modes in the Gravitational-wave Background},
  Astrophys. J. Lett. 923~(2) (2021) L22.
\newblock \href {http://arxiv.org/abs/2109.14706} {\path{arXiv:2109.14706}},
  \href {http://dx.doi.org/10.3847/2041-8213/ac401c}
  {\path{doi:10.3847/2041-8213/ac401c}}.

\bibitem{Chen:2021wdo}
Z.-C. Chen, C.~Yuan, Q.-G. Huang, {Non-tensorial gravitational wave background
  in NANOGrav 12.5-year data set}, Sci. China Phys. Mech. Astron. 64~(12)
  (2021) 120412.
\newblock \href {http://arxiv.org/abs/2101.06869} {\path{arXiv:2101.06869}},
  \href {http://dx.doi.org/10.1007/s11433-021-1797-y}
  {\path{doi:10.1007/s11433-021-1797-y}}.

\bibitem{Chen:2021ncc}
Z.-C. Chen, Y.-M. Wu, Q.-G. Huang, {Searching for Isotropic Stochastic
  Gravitational-Wave Background in the International Pulsar Timing Array Second
  Data Release}\href {http://arxiv.org/abs/2109.00296}
  {\path{arXiv:2109.00296}}.

\bibitem{Dai:2012bc}
L.~Dai, M.~Kamionkowski, D.~Jeong, {Total Angular Momentum Waves for Scalar,
  Vector, and Tensor Fields}, Phys. Rev. D 86 (2012) 125013.
\newblock \href {http://arxiv.org/abs/1209.0761} {\path{arXiv:1209.0761}},
  \href {http://dx.doi.org/10.1103/PhysRevD.86.125013}
  {\path{doi:10.1103/PhysRevD.86.125013}}.

\bibitem{Qin:2018yhy}
W.~Qin, K.~K. Boddy, M.~Kamionkowski, L.~Dai, {Pulsar-timing arrays,
  astrometry, and gravitational waves}, Phys. Rev. D 99~(6) (2019) 063002.
\newblock \href {http://arxiv.org/abs/1810.02369} {\path{arXiv:1810.02369}},
  \href {http://dx.doi.org/10.1103/PhysRevD.99.063002}
  {\path{doi:10.1103/PhysRevD.99.063002}}.

\bibitem{Qin:2020hfy}
W.~Qin, K.~K. Boddy, M.~Kamionkowski, {Subluminal stochastic gravitational
  waves in pulsar-timing arrays and astrometry}, Phys. Rev. D 103~(2) (2021)
  024045.
\newblock \href {http://arxiv.org/abs/2007.11009} {\path{arXiv:2007.11009}},
  \href {http://dx.doi.org/10.1103/PhysRevD.103.024045}
  {\path{doi:10.1103/PhysRevD.103.024045}}.

\bibitem{Deffayet:2009wt}
C.~Deffayet, G.~Esposito-Farese, A.~Vikman, {Covariant Galileon}, Phys. Rev. D
  79 (2009) 084003.
\newblock \href {http://arxiv.org/abs/0901.1314} {\path{arXiv:0901.1314}},
  \href {http://dx.doi.org/10.1103/PhysRevD.79.084003}
  {\path{doi:10.1103/PhysRevD.79.084003}}.

\bibitem{DeFelice:2010pv}
A.~De~Felice, S.~Tsujikawa, {Cosmology of a covariant Galileon field}, Phys.
  Rev. Lett. 105 (2010) 111301.
\newblock \href {http://arxiv.org/abs/1007.2700} {\path{arXiv:1007.2700}},
  \href {http://dx.doi.org/10.1103/PhysRevLett.105.111301}
  {\path{doi:10.1103/PhysRevLett.105.111301}}.

\bibitem{deRham:2012az}
C.~de~Rham, {Galileons in the Sky}, Comptes Rendus Physique 13 (2012) 666--681.
\newblock \href {http://arxiv.org/abs/1204.5492} {\path{arXiv:1204.5492}},
  \href {http://dx.doi.org/10.1016/j.crhy.2012.04.006}
  {\path{doi:10.1016/j.crhy.2012.04.006}}.

\bibitem{Joyce:2014kja}
A.~Joyce, B.~Jain, J.~Khoury, M.~Trodden, {Beyond the Cosmological Standard
  Model}, Phys. Rept. 568 (2015) 1--98.
\newblock \href {http://arxiv.org/abs/1407.0059} {\path{arXiv:1407.0059}},
  \href {http://dx.doi.org/10.1016/j.physrep.2014.12.002}
  {\path{doi:10.1016/j.physrep.2014.12.002}}.

\bibitem{Brax:2011sv}
P.~Brax, C.~Burrage, A.-C. Davis, {Laboratory Tests of the Galileon}, JCAP 09
  (2011) 020.
\newblock \href {http://arxiv.org/abs/1106.1573} {\path{arXiv:1106.1573}},
  \href {http://dx.doi.org/10.1088/1475-7516/2011/09/020}
  {\path{doi:10.1088/1475-7516/2011/09/020}}.

\bibitem{Ali:2012cv}
A.~Ali, R.~Gannouji, M.~W. Hossain, M.~Sami, {Light mass galileons:
  Cosmological dynamics, mass screening and observational constraints}, Phys.
  Lett. B 718 (2012) 5--14.
\newblock \href {http://arxiv.org/abs/1207.3959} {\path{arXiv:1207.3959}},
  \href {http://dx.doi.org/10.1016/j.physletb.2012.10.009}
  {\path{doi:10.1016/j.physletb.2012.10.009}}.

\bibitem{Andrews:2013qva}
M.~Andrews, Y.-Z. Chu, M.~Trodden, {Galileon forces in the Solar System}, Phys.
  Rev. D 88 (2013) 084028.
\newblock \href {http://arxiv.org/abs/1305.2194} {\path{arXiv:1305.2194}},
  \href {http://dx.doi.org/10.1103/PhysRevD.88.084028}
  {\path{doi:10.1103/PhysRevD.88.084028}}.

\bibitem{Appleby:2018yci}
S.~Appleby, E.~V. Linder, {The Well-Tempered Cosmological Constant}, JCAP 07
  (2018) 034.
\newblock \href {http://arxiv.org/abs/1805.00470} {\path{arXiv:1805.00470}},
  \href {http://dx.doi.org/10.1088/1475-7516/2018/07/034}
  {\path{doi:10.1088/1475-7516/2018/07/034}}.

\bibitem{Appleby:2022bxp}
S.~Appleby, R.~C. Bernardo, {Tadpole Cosmology: Self Tuning Without
  Degeneracy}, JCAP 07 (2022) 035.
\newblock \href {http://arxiv.org/abs/2202.08672} {\path{arXiv:2202.08672}},
  \href {http://dx.doi.org/10.1088/1475-7516/2022/07/035}
  {\path{doi:10.1088/1475-7516/2022/07/035}}.

\bibitem{Brax:2014vla}
P.~Brax, C.~Burrage, A.-C. Davis, G.~Gubitosi, {Cosmological Tests of Coupled
  Galileons}, JCAP 03 (2015) 028.
\newblock \href {http://arxiv.org/abs/1411.7621} {\path{arXiv:1411.7621}},
  \href {http://dx.doi.org/10.1088/1475-7516/2015/03/028}
  {\path{doi:10.1088/1475-7516/2015/03/028}}.

\bibitem{Burrage:2010cu}
C.~Burrage, C.~de~Rham, D.~Seery, A.~J. Tolley, {Galileon inflation}, JCAP 01
  (2011) 014.
\newblock \href {http://arxiv.org/abs/1009.2497} {\path{arXiv:1009.2497}},
  \href {http://dx.doi.org/10.1088/1475-7516/2011/01/014}
  {\path{doi:10.1088/1475-7516/2011/01/014}}.

\bibitem{LIGOScientific:2017vwq}
B.~P. Abbott, et~al., {GW170817: Observation of Gravitational Waves from a
  Binary Neutron Star Inspiral}, Phys. Rev. Lett. 119~(16) (2017) 161101.
\newblock \href {http://arxiv.org/abs/1710.05832} {\path{arXiv:1710.05832}},
  \href {http://dx.doi.org/10.1103/PhysRevLett.119.161101}
  {\path{doi:10.1103/PhysRevLett.119.161101}}.

\bibitem{LIGOScientific:2021sio}
R.~Abbott, et~al., {Tests of General Relativity with GWTC-3}\href
  {http://arxiv.org/abs/2112.06861} {\path{arXiv:2112.06861}}.

\bibitem{Jenet:2014bea}
F.~A. Jenet, J.~D. Romano, {Understanding the gravitational-wave Hellings and
  Downs curve for pulsar timing arrays in terms of sound and electromagnetic
  waves}, Am. J. Phys. 83 (2015) 635.
\newblock \href {http://arxiv.org/abs/1412.1142} {\path{arXiv:1412.1142}},
  \href {http://dx.doi.org/10.1119/1.4916358} {\path{doi:10.1119/1.4916358}}.

\bibitem{Ng:2021waj}
K.-W. Ng, {Redshift-space fluctuations in stochastic gravitational wave
  background}, Phys. Rev. D 106~(4) (2022) 043505.
\newblock \href {http://arxiv.org/abs/2106.12843} {\path{arXiv:2106.12843}},
  \href {http://dx.doi.org/10.1103/PhysRevD.106.043505}
  {\path{doi:10.1103/PhysRevD.106.043505}}.

\bibitem{Chu:2021krj}
Y.-K. Chu, G.-C. Liu, K.-W. Ng, {Observation of a polarized stochastic
  gravitational-wave background in pulsar-timing-array experiments}, Phys. Rev.
  D 104~(12) (2021) 124018.
\newblock \href {http://arxiv.org/abs/2107.00536} {\path{arXiv:2107.00536}},
  \href {http://dx.doi.org/10.1103/PhysRevD.104.124018}
  {\path{doi:10.1103/PhysRevD.104.124018}}.

\bibitem{Trotta:2008qt}
R.~Trotta, {Bayes in the sky: Bayesian inference and model selection in
  cosmology}, Contemp. Phys. 49 (2008) 71--104.
\newblock \href {http://arxiv.org/abs/0803.4089} {\path{arXiv:0803.4089}},
  \href {http://dx.doi.org/10.1080/00107510802066753}
  {\path{doi:10.1080/00107510802066753}}.

\bibitem{Lewis:2019xzd}
A.~Lewis, \href{https://getdist.readthedocs.io}{{GetDist: a Python package for
  analysing Monte Carlo samples}}\href {http://arxiv.org/abs/1910.13970}
  {\path{arXiv:1910.13970}}.
\newline\urlprefix\url{https://getdist.readthedocs.io}

\bibitem{Torrado:2020dgo}
J.~Torrado, A.~Lewis, {Cobaya: Code for Bayesian Analysis of hierarchical
  physical models}, JCAP 05 (2021) 057.
\newblock \href {http://arxiv.org/abs/2005.05290} {\path{arXiv:2005.05290}},
  \href {http://dx.doi.org/10.1088/1475-7516/2021/05/057}
  {\path{doi:10.1088/1475-7516/2021/05/057}}.

\bibitem{Liddle:2007fy}
A.~R. Liddle, {Information criteria for astrophysical model selection}, Mon.
  Not. Roy. Astron. Soc. 377 (2007) L74--L78.
\newblock \href {http://arxiv.org/abs/astro-ph/0701113}
  {\path{arXiv:astro-ph/0701113}}, \href
  {http://dx.doi.org/10.1111/j.1745-3933.2007.00306.x}
  {\path{doi:10.1111/j.1745-3933.2007.00306.x}}.

\bibitem{Chamberlin:2011ev}
S.~J. Chamberlin, X.~Siemens, {Stochastic backgrounds in alternative theories
  of gravity: overlap reduction functions for pulsar timing arrays}, Phys. Rev.
  D 85 (2012) 082001.
\newblock \href {http://arxiv.org/abs/1111.5661} {\path{arXiv:1111.5661}},
  \href {http://dx.doi.org/10.1103/PhysRevD.85.082001}
  {\path{doi:10.1103/PhysRevD.85.082001}}.

\bibitem{rbgithub}
R.~Bernardo,
  \href{https://doi.org/10.5281/zenodo.6973318}{reggiebernardo/galileon\_pta:
  zeroth} (Aug. 2022).
\newblock \href {http://dx.doi.org/10.5281/zenodo.6973317}
  {\path{doi:10.5281/zenodo.6973317}}.
\newline\urlprefix\url{https://doi.org/10.5281/zenodo.6973318}

\end{thebibliography}


\clearpage
\onecolumn

\setcounter{equation}{0}
\setcounter{page}{1}
\setcounter{figure}{0}

\section*{\LARGE\bfseries Looking out for the Galileon in the nanohertz gravitational wave sky: \\ Supplementary notes}
\label{sec:supp_notes}

\subsection{Covariant field equations}
\label{sec:field_equations}

The variation of covariant Galileon action with respect to the metric leads to the modified Einstein equation
\be
\begin{split}
\mpl G_{bc} \left( 1 + \alpha \dfrac{\phi}{M_\text{pl}} \right) 
& + M_\text{pl} \alpha \left( - \nabla_b \nabla_c \phi + g_{bc} \Box \phi \right) + \Lambda g_{bc} \\
& + \lambda^3 g_{bc} \phi + \dfrac{\mu^2 \phi^2}{2} g_{bc} - \left( \nabla_b \phi \right) \left( \nabla_c \phi \right) + \dfrac{\left( \partial \phi \right)^2}{2} g_{bc} \\
& + \dfrac{1}{\kappa^3} \left( -(\nabla_b \phi) (\nabla_c \phi) \Box \phi + 2 (\nabla^d \phi) (\nabla_d \nabla_{(b} \phi) (\nabla_{c)} \phi) - g_{bc} 
(\nabla^d \phi)(\nabla^e \phi) \nabla_d \nabla_e \phi \right) = T_{bc} \,,
\end{split}
\ee
where $T_{bc}$ is the matter stress-energy tensor and index symmetrization is defined as $V_{(ab)} = (V_{ab} + V_{ba})/2$. On the other hand, the variation with respect to the scalar field gives the modified Klein-Gordon equation
\be
\begin{split}
    \Box \phi - \mu^2 \phi + \dfrac{1}{\kappa^3} \left( \left( \Box \phi \right)^2 - \left( \nabla_b \nabla_c \phi \right) \left(\nabla^b \nabla^c \phi\right) - R_{bc} \left( \nabla^b \phi \right) \left( \nabla^c \phi \right) \right) = \lambda^3 - \dfrac{M_\text{pl} \alpha R}{2} \,.
\end{split}
\ee

\subsection{Linearized field equations}
\label{sec:linearized_eqs}

We study the perturbations of the covariant Galileon about its static vacuum $R_{ab} = 0$.

We work in the synchronous gauge:
\be
ds^2 = -dt^2 + \left( \delta_{AB} - 2 \psi \delta_{AB} + 2 D_A D_B E + 2 D_{(A} E_{B)} + 2 E_{AB} \right) dx^A dx^B \,,
\ee
where $\psi$ and $E$ are scalars, $E_A$ is a transverse vector ($D_A E^A = 0$), and $E_{AB}$ is a transverse-traceless tensor ($D^A E_{AB} = 0$ and $\delta^{AB} E_{AB} = 0$). $D$ is the spatial covariant derivative operator. We consider the scalar field perturbation
\be
\phi = \varphi + \delta \phi \,.
\ee
A simple counting of components, $\psi (+1), E (+1), E_B (+2), E_{AB} (+2)$, shows that the gauge is now fully fixed. We turn off the vector modes ($E_B = 0$) as they are nondynamical and simply diluted by the expansion in the context of the Galileon cosmology.

We start with the scalar perturbations $\psi$, $E$, and $\delta \phi$. The Hamiltonian constraint, or rather the time-time component of the linearized modified Einstein equation, can be simplified to
\be
D^2 \left( -\dfrac{M_\text{pl} \alpha}{2} \delta \phi + \mpl \left( 1 - \dfrac{\alpha \lambda^3}{M_\text{pl} \mu^2} \right) \psi \right) = 0 \,.
\ee
Obviously, this is a Poisson equation $D^2 \Psi = 0$ for a scalar $\Psi[\psi, \delta \phi]$. This relates the scalar potentials $\psi$ and $\delta \phi$ up to a constant which we can set to zero by taking the perturbations to vanish at infinity. We thus identify
\be
\delta \phi = \dfrac{2 M_\text{pl}}{\alpha} \left( 1 - \dfrac{\alpha \lambda^3}{M_\text{pl} \mu^2} \right) \psi \,
\ee
which we use to eliminate $\delta \phi$ in the upcoming relations. The momentum constraint (or the time-space component) turns out to be $D_B \Psi = 0$ which leads to the same relation between the scalar potentials as did the Hamiltonian constraint. Moving to the spatial component, we find that the spatial trace gives
\be
D^2 \left( \left( 1 - \dfrac{\alpha \lambda^3}{ M_\text{pl} \mu^2 } \right) \left( \psi - \ddot{E} \right) \right) = 0 \,,
\ee
where a dot denotes a time derivative. This is once again a Poisson equation and allows us to write down
\be
\ddot{E} = \psi \,.
\ee
Now, we take a look at the linearized scalar field equation. By using the above relations, we are able to simplify this to
\be
\left( 1 + \dfrac{3 \alpha^2}{2} - \dfrac{\alpha \lambda^3}{ M_\text{pl} \mu^2} \right) \left( D^2 \psi - \ddot{\psi} \right) - \mu^2 \left( 1 - \dfrac{\alpha\lambda^3}{M_\text{pl}\mu^2} \right) \psi = 0 \,.
\ee
This shows that the scalar perturbations are controlled by three parameters: $\mu$, $\alpha$ and $\lambda^3/ \left( M_\text{pl} \mu^2 \right)$.

The propagating scalar modes of the Galileon therefore satisfy a massive dispersion relation
\be
\omega^2 = k^2 + m_\text{eff}^2
\ee
where $\omega$ and $\vec{k} = k \hat{k}$ are the frequency and wave number, respectively, and the effective mass $m_\text{eff}$ is given by
\be
m_\text{eff}^2 = \mu^2 \dfrac{\left( 1 - \alpha \lambda^3 / (M_\text{pl} \mu^2) \right) }{ \left( 1 + (3\alpha^2/2) - \alpha \lambda^3/(M_\text{pl} \mu^2) \right)} \, .
\ee
The conformal coupling changes the effective mass of the scalar modes. Such massive dispersion relation leads to a group velocity $v = d\omega/dk = k/\omega$ that is reciprocal of the phase velocity $v_\text{ph} = 1/v = \omega/k$. Combining the above results, we find that the scalar perturbations in the metric can be written as
\be
ds^2 = - dt^2 + \left( \delta_{AB} - 2 \psi \left( \delta_{AB} - v^2 \dfrac{k_A k_B}{k^2} \right) \right) dx^A dx^B \,,
\ee
where $v$ is the group velocity. This brings in scalar transverse and longitudinal polarizations as we now show.

We consider a wave propagating along the $z$-axis ($k_A = (0, 0, k)$). In this case, we write down
\be
h_{AB} = 2\psi \left( \delta_{AB} - v^2 \dfrac{k_A k_B}{k^2} \right) \propto 
\left(
\begin{array}{ccc}
    1 & & \\
     & 1 & \\
     & & 1 - v^2
\end{array}
\right) \,.
\ee
The nonzero $z$ component reveals this is a mixture of transverse and longitudinal modes. We understand this by expressing the gravitational wave as a sum of various polarizations. To be concrete, we take the polarization tensors
\be
\varepsilon_{AB}^{\text{ST}} = \left(
\begin{array}{ccc}
    1 & & \\
     & 1 & \\
     & & 0
\end{array}
\right)
\ee
and
\be
\varepsilon_{AB}^{\text{SL}} = \sqrt{2} \left(
\begin{array}{ccc}
    0 & & \\
     & 0 & \\
     & & 1
\end{array}
\right)
\ee
normalized such that $\varepsilon^{P, AB} \varepsilon^{P'}_{AB} = 2 \delta^{PP'}$ where $P = \text{ST}, \text{SL}$ stands for `scalar transverse' and `scalar longitudinal' polarizations. Then, we decompose the metric perturbation as
\be
h_{AB} \propto \left( \varepsilon_{AB}^{\text{ST}} + \dfrac{1-v^2}{\sqrt{2}} \varepsilon_{AB}^{\text{SL}} \right) \tilde{\psi}\left(k\right) e^{i(k z - \omega t)}
\ee
and from which identify the ratio of the scalar transverse to scalar longitudinal amplitudes as $\left(1 - v^2\right)/\sqrt{2} = m_\text{eff}^2/\left(\sqrt{2}\omega^2\right)$.

\subsection{Autocorrelation using the real space formalism}
\label{subsec:autocorrelation_rsf}

We summarize the main results for the autocorrelation function $\Gamma_{aa}^A$ used in the paper which were obtained using the real space formalism.

As a two sphere integral, the overlap reduction function for a polarization $A$ of velocity $v$ can be shown to be
\begin{equation}
\label{eq:orf_realspace}
    \Gamma_{ab}^A\left( \zeta, f D_i \right) = \int_{S^2} d \hat{k} \ U_a \left(f D_a, \hat{k} \right) U_b^*\left(f D_b, \hat{k} \right) F_a^A \left( \hat{k} \right) F_b^A \left( \hat{k} \right) \,,
\end{equation}
where the antenna pattern functions $F_a^A\left(\hat{k}\right)$ are given by 
\begin{equation}
\label{eq:antenna_functions}
    F_a^A\left(\hat{k}\right) = \dfrac{\left( \hat{e}_a^i \otimes \hat{e}_a^j \right) \cdot \varepsilon_{ij}^A \left( \hat{k} \right)}{2\left( 1 + v \hat{k} \cdot \hat{e}_a \right)}
\end{equation}
and
\begin{equation}
\label{eq:Uadef}
    U_a\left(f D_a, \hat{k}\right) = 1 - e^{2 \pi i f D_a \left( 1 + v \hat{k} \cdot \hat{e}_a \right)} \,.
\end{equation}
The autocorrelation can be obtained by the correlation of a pulsar with itself, or rather by setting $a = b$ in the overlap reduction function. 

We begin with the tensor modes. This requires the antenna pattern functions for the tensor $+$ and $\times$ modes which are given by
\begin{equation}
    F^+_a \left( \hat{k} = (\theta, \phi) \right) = \dfrac{\cos(2\phi) \sin^2\theta}{2 \left( 1 + v \cos \theta \right)}
\end{equation}
and
\begin{equation}
    F^\times_a \left( \hat{k} = (\theta, \phi) \right) = \dfrac{\sin(2\phi) \sin^2\theta}{2 \left( 1 + v \cos \theta \right)} \,.
\end{equation}
The tensor autocorrelation is then given by the integral
\begin{equation}
    \Gamma_{aa}^\text{T} = \int_0^\pi d\theta \left( \frac{2 \pi  \sin ^5 \theta  \sin ^2(\pi  fD (1 + v \cos \theta))}{(1 + v \cos \theta )^2} \right) \,. 
\end{equation}
On the other hand, the antenna pattern functions for the scalar transverse and longitudinal modes are
\begin{equation}
    F^\text{ST}_a \left( \hat{k} = (\theta, \phi) \right) = \dfrac{\sin^2 \theta}{2 \left( 1 + v \cos \theta \right)}
\end{equation}
and
\begin{equation}
    F^\text{SL}_a \left( \hat{k} = (\theta, \phi) \right) = \dfrac{\cos^2\theta}{\sqrt{2} \left( 1 + v \cos \theta \right)} \,,
\end{equation}
respectively. The scalar autocorrelation function reduces to the integrals
\begin{equation}
    \Gamma_{aa}^\text{ST} = \int_0^\pi d\theta \left( \frac{2 \pi  \sin ^5 \theta \sin ^2 ( \pi fD (1 + v \cos \theta ))}{(1 + v \cos \theta)^2} \right) 
\end{equation}
and
\begin{equation}
    \Gamma_{aa}^\text{SL} = \int_0^\pi d\theta \left( \frac{4 \pi  \sin \theta \cos ^4 \theta \sin ^2 \left(\pi fD (1 + v \cos \theta ) \right)}{(1 + v \cos \theta )^2} \right) \,. 
\end{equation}
It is interesting that $\Gamma_{aa}^\text{ST}$ coincides with the transverse tensor $\Gamma_{aa}^\text{T}$.

\subsection{Scalar power spectra phenomenology}
\label{sec:scalar_ps_phenom}

\begin{figure}[h!]
\center
	\subfigure[ \, scalar transverse, $v = 99/100$ ]{
		\includegraphics[width = 0.45 \textwidth]{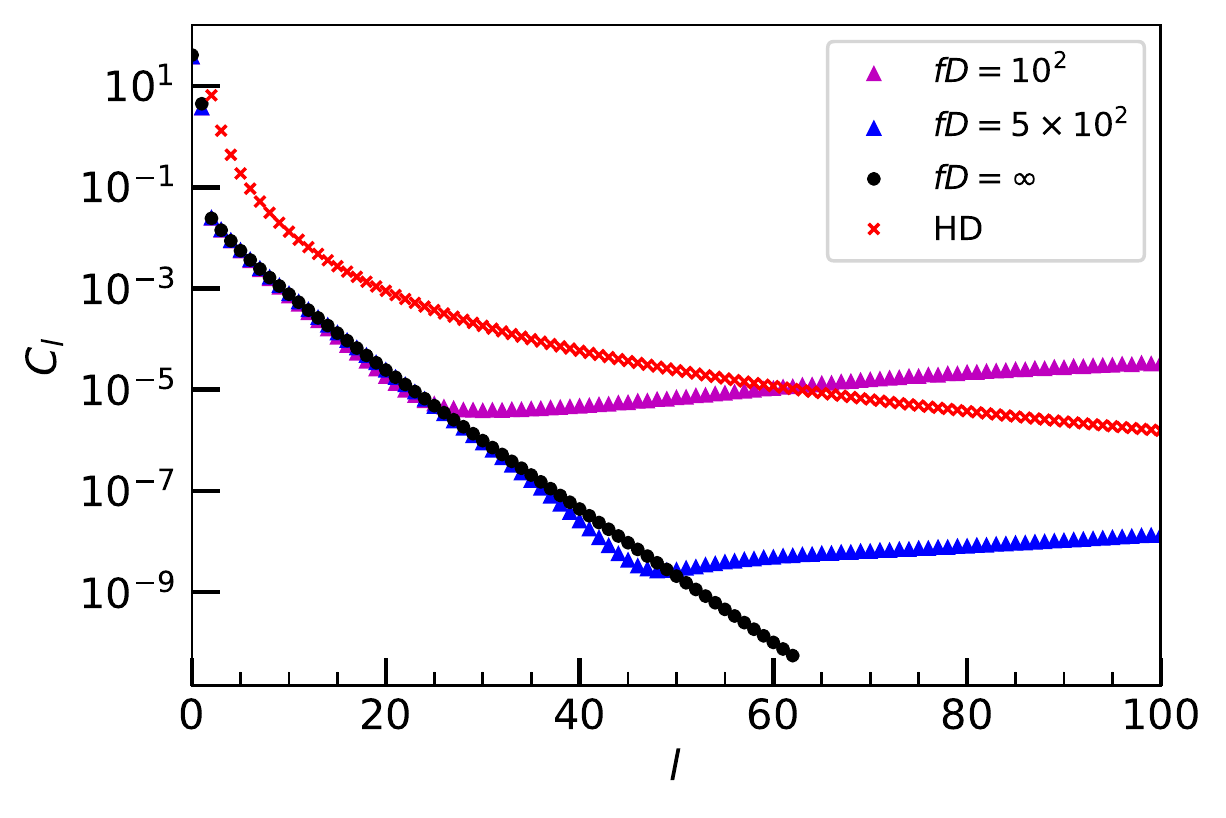}
		}
	\subfigure[ \, scalar transverse, $v = 99/100$ ]{
		\includegraphics[width = 0.45 \textwidth]{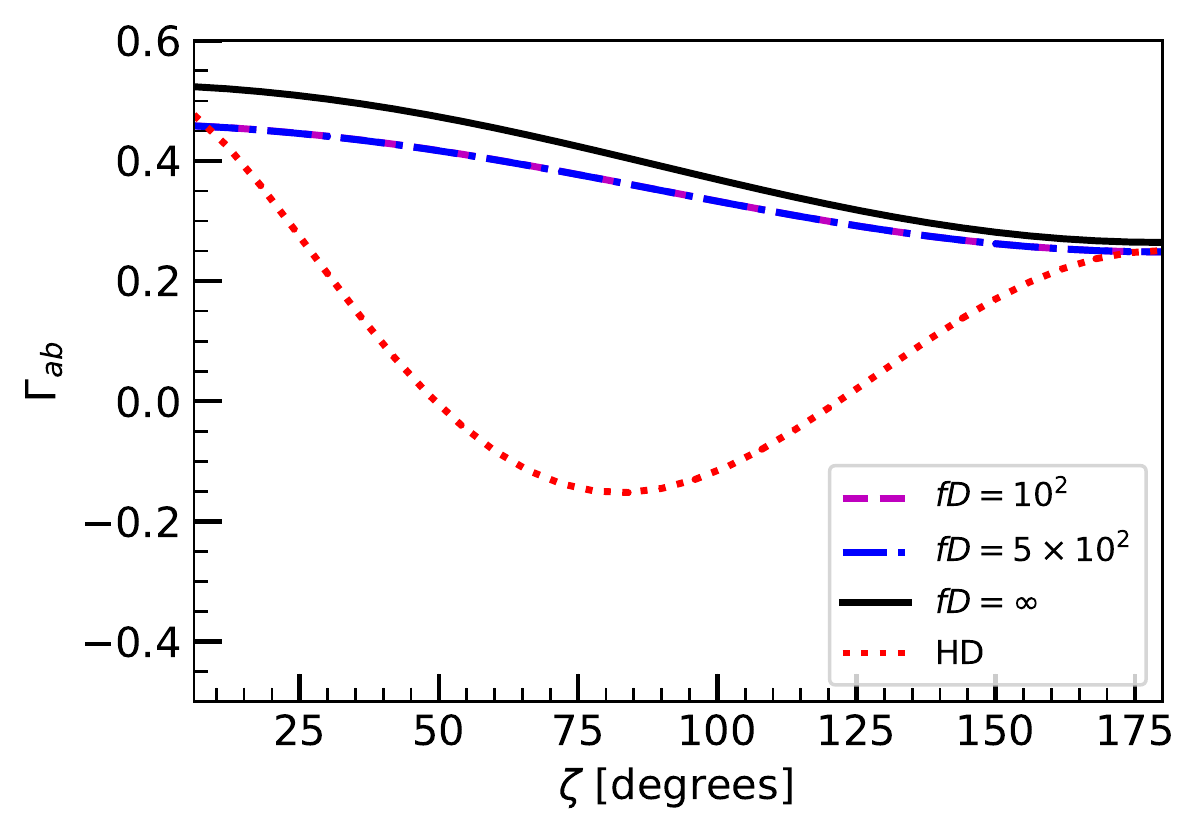}
		}
	\subfigure[ \, scalar transverse, $v = 1/2$ ]{
		\includegraphics[width = 0.45 \textwidth]{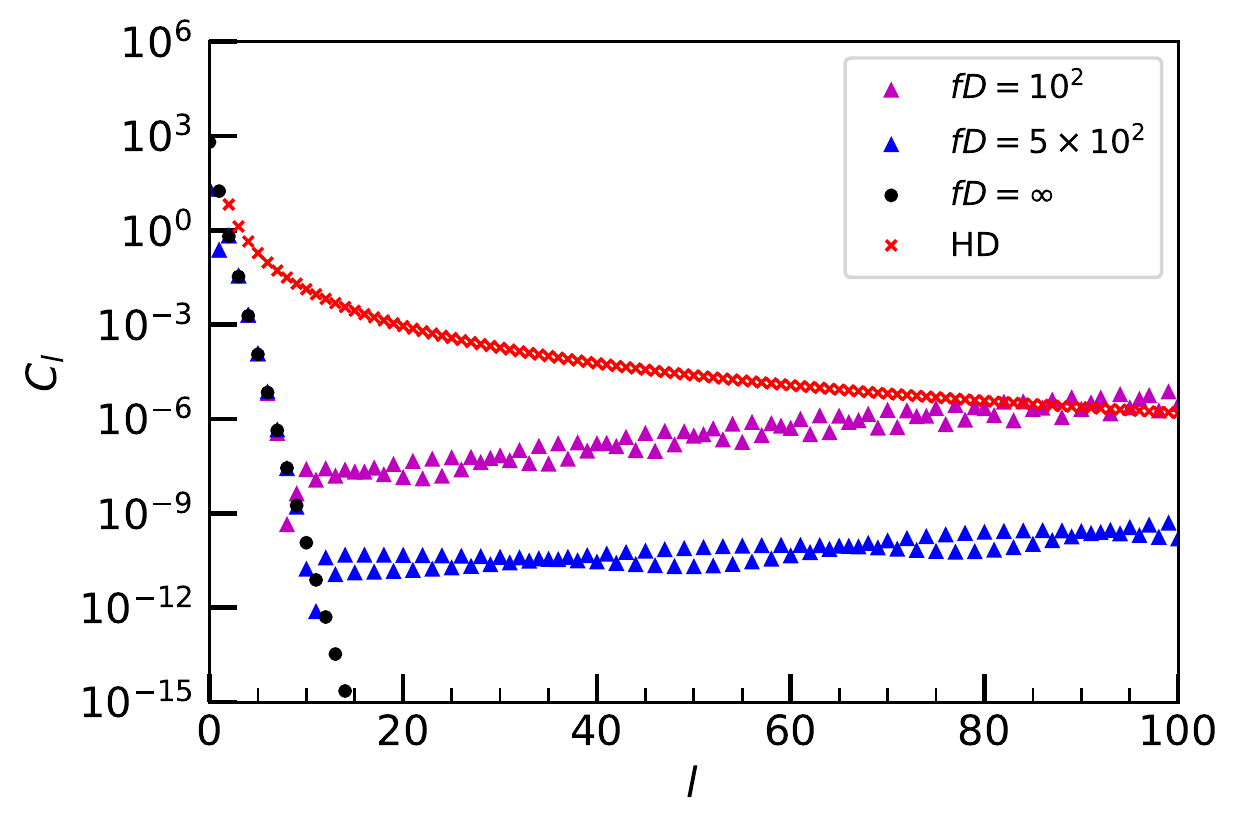}
		}
	\subfigure[ \, scalar transverse, $v = 1/2$ ]{
		\includegraphics[width = 0.45 \textwidth]{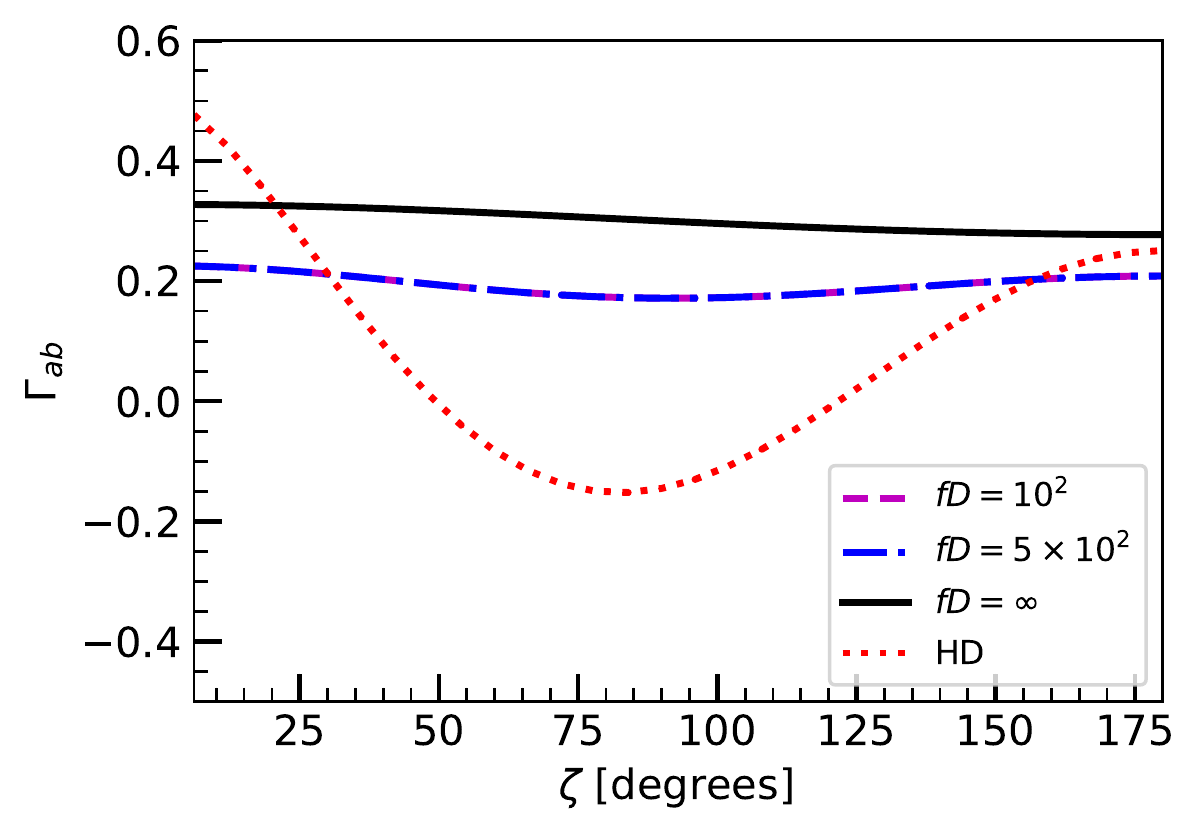}
		}
	\subfigure[ \, scalar transverse, $v = 1/100$ ]{
		\includegraphics[width = 0.45 \textwidth]{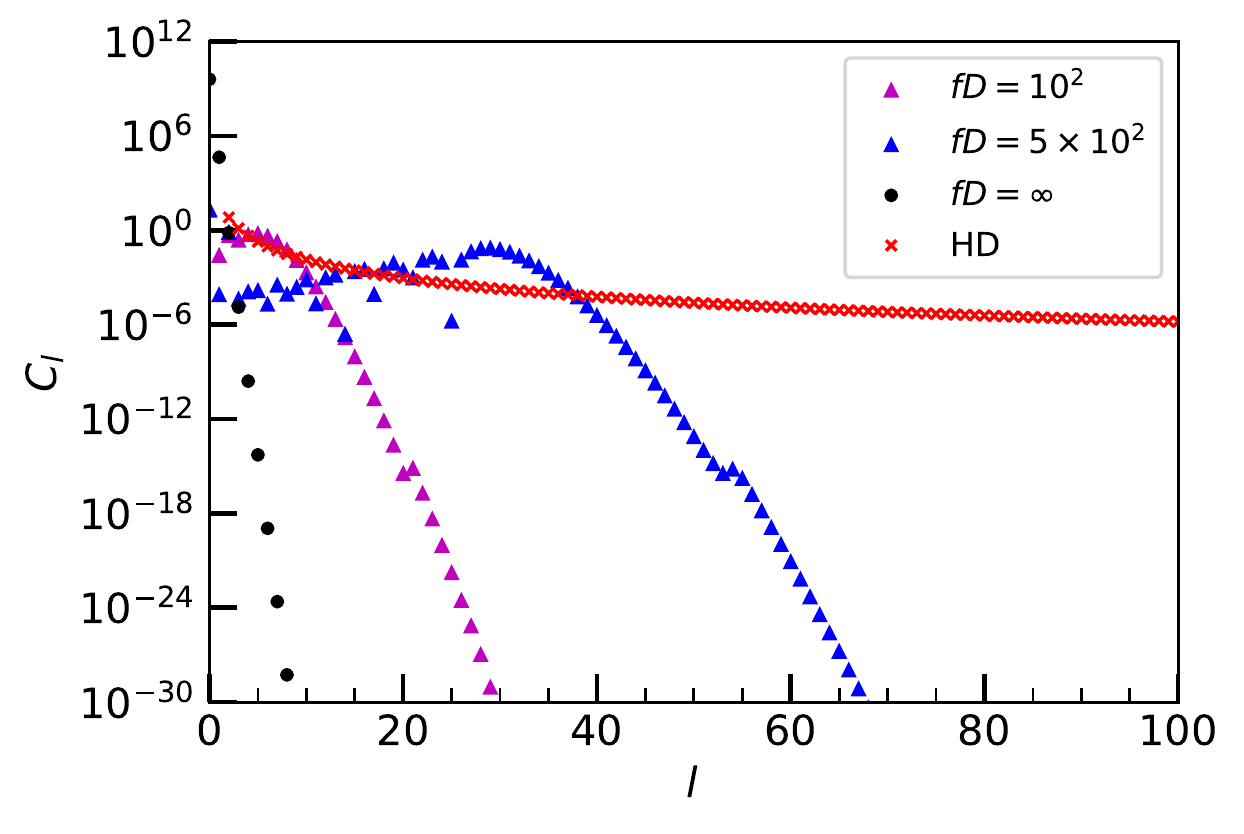}
		}
	\subfigure[ \, scalar transverse, $v = 1/100$ ]{
		\includegraphics[width = 0.45 \textwidth]{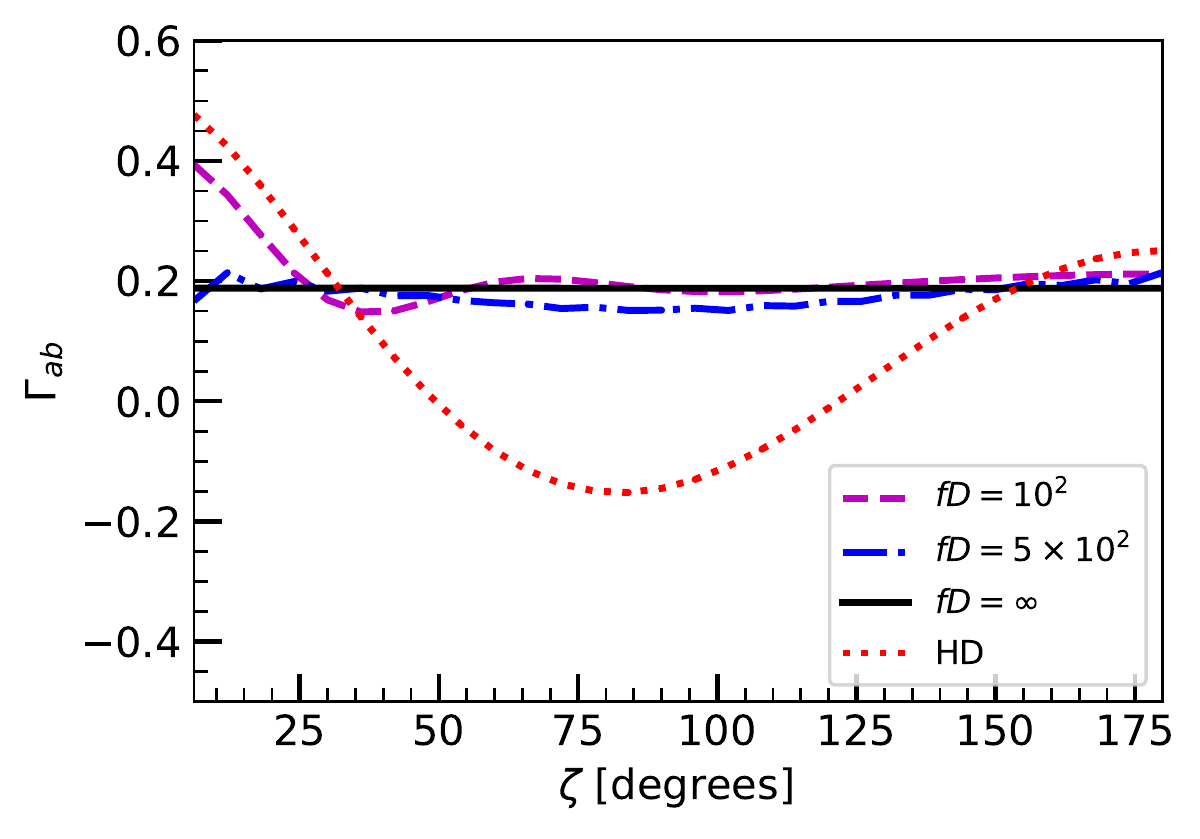}
		}
\caption{The isotropic power spectra multipoles $C_l$ and the overlap reduction function $\Gamma_{ab}(\zeta)$ of the scalar transverse polarization with group velocities $v = 99/100$, $v = 1/2$, and $v = 1/100$. The overlap reduction functions were constructed with only the first thirty multipoles and are normalized with respect to the Hellings-Downs (HD) correlation.}
\label{fig:ClST}
\end{figure}

We present the power spectra and resulting overlap reduction function individually for each of the scalar polarizations. Figure \ref{fig:ClST} shows this for the scalar transverse polarization with $v = 99/100$ (near luminal), $v = 1/2$ (half the speed of light), and $v = 10^{-2}$ (near static) at various pulsar distances. We include the Hellings-Downs signal in the plots for reference.

First off, this reveals an important physical difference between the finite and infinite pulsar distance cases. As reflected in each power spectrum, the multipolar power in the infinite distance limit drops continuously at small angular scales, or rather for increasing multipole index $l$. This physically corresponds to missing about half the overall power that should be for pulsar pairs nearly at the same location in the sky. On the other hand, this missing small scale power in the infinite distance limit is in fact captured by the finite distance power spectrum. The infinite distance limit implies that the correlations vanish for pulsars that are infinitesimally separated in the sky. However, in a real setting, pulsars are separated at a finite line-of-sight distance from the observer (`Us'). The results tell that correlations at small angular separations are strengthened for nearby pulsars.

Figure \ref{fig:ClST}(a) shows that at some $l \sim 20 - 50$ with nearly luminal scalar ($v \sim 1$), the power spectra ceases to drop and instead sustains a slow increase. A similar partial decay was recognized in the transverse-traceless tensor polarization for finite pulsar distances. The corresponding overlap reduction function for the near luminal scalar is shown in Figure \ref{fig:ClST}(b) where only the first thirty multipoles were considered, or rather that correlations for pulsar pairs separated by more than six degrees can be trusted. This shows the scalar transverse signal resembling what looks more like a dipole, which can be explained by its power spectrum which is dominated by the monopole ($l = 0$) and dipole ($l = 1$). The overlap reduction function for the finite distance cases are also noticeably indistinguishable since their first few power spectra multipoles, corresponding to large angles, are the same.

Figures \ref{fig:ClST}(c-d) show the multipoles and the overlap reduction functions when the scalar modes propagate at half the speed of light. In this case, we find that the power spectra low multipoles ($l \lesssim 10$) feature a sharper drop in the power. This is reflected in the overlap reduction function which tended to flatter values as compared with the luminal scalar case, since now the power is dominated by the monopole, dipole, and the quadrupole. Notice, however, that the overlap reduction function for the infinite distance limit looks like a dipole while that for the finite distance cases have a more pronounced quadrupolar profile. The overlap reduction functions for the finite pulsar distance cases for large angular separations relevant for pulsar timing array in Figure \ref{fig:ClST}(d) are visually indistinguishable as displayed by their multipoles (Figure \ref{fig:ClST}(c)). Further, the power continues to drop sharply at higher multipoles in the infinite distance limit, compared with the finite distance cases which contain a small scale power.

When the velocity is decreased further to near nonrelativistic speeds, the power spectrum profile of the infinite distance limit drops even sharper, so that the monopole becomes practically the relevant component. Most interestingly, in this velocity limit, the finite distance cases feature a beyond quadrupolar peak ($l_\text{peak} \neq 0, 1, 2$) in the power spectrum, which presents as an oscillation in the overlap reduction function (i.e., oscillation starts at $\zeta \sim 180^\circ/l_\text{peak}$). Figures \ref{fig:ClST}(e-f) show this with $v = 10^{-2}$. The large angle multipolar profile, with $l_\text{peak}$, now allows the finite distance cases to be distinguishable through their overlap reduction functions. This also caters a drastic difference in this observable between the infinite distance limit (i.e., a horizontal line) and the finite distance cases which retain physically interesting angular profiles. The finite distance cases not only capture small scale power missed in the infinite distance limit, but also present oscillations which represent pulsars as astrophysical objects.

\begin{figure}[h!]
\center
	\subfigure[ \, scalar longitudinal, $v = 99/100$ ]{
		\includegraphics[width = 0.45 \textwidth]{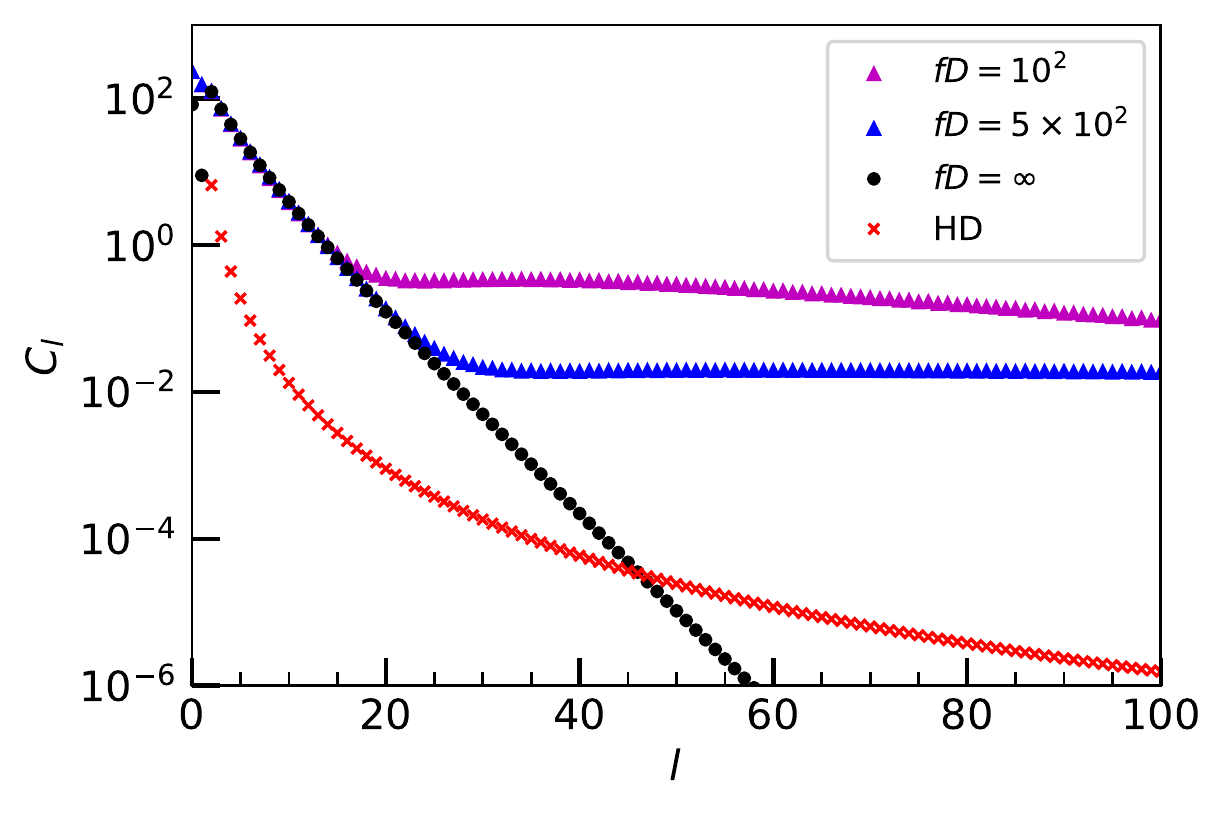}
		}
	\subfigure[ \, scalar longitudinal, $v = 99/100$ ]{
		\includegraphics[width = 0.45 \textwidth]{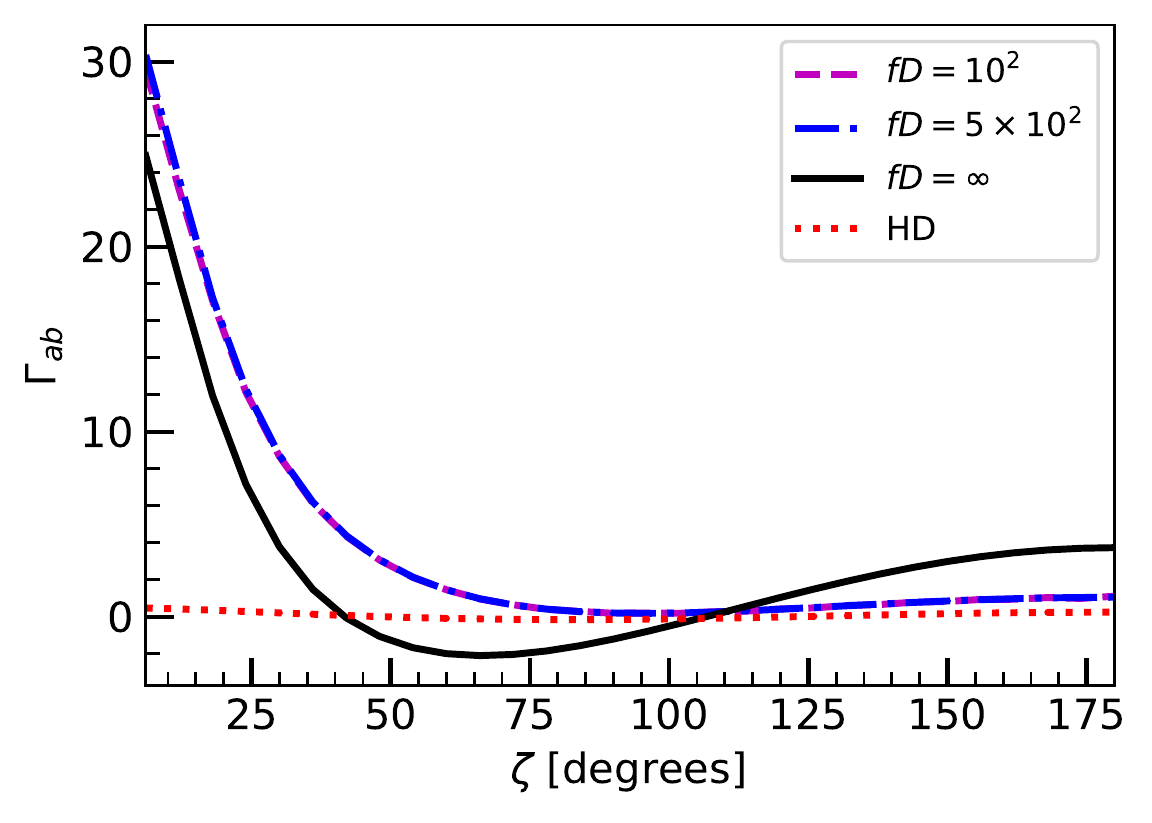}
		}
	\subfigure[ \, scalar longitudinal, $v = 1/2$ ]{
		\includegraphics[width = 0.45 \textwidth]{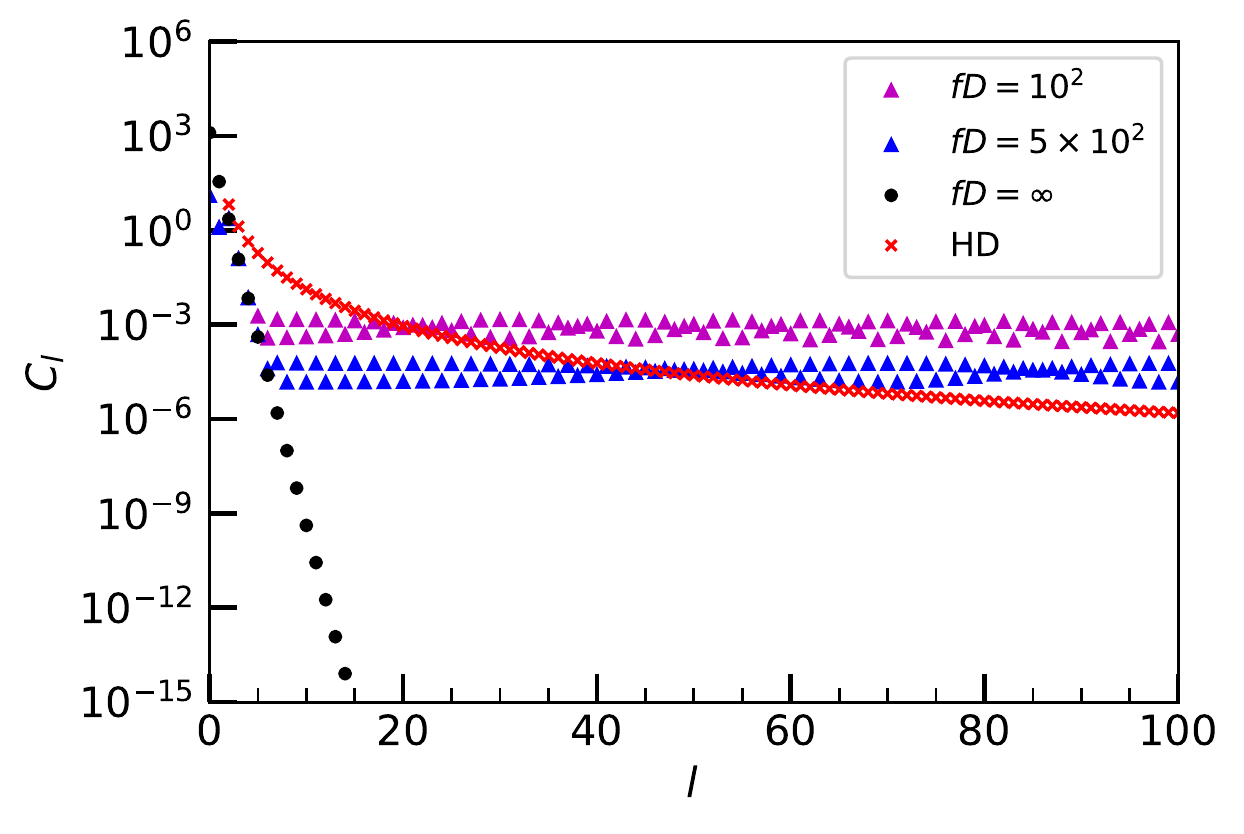}
		}
	\subfigure[ \, scalar longitudinal, $v = 1/2$ ]{
		\includegraphics[width = 0.45 \textwidth]{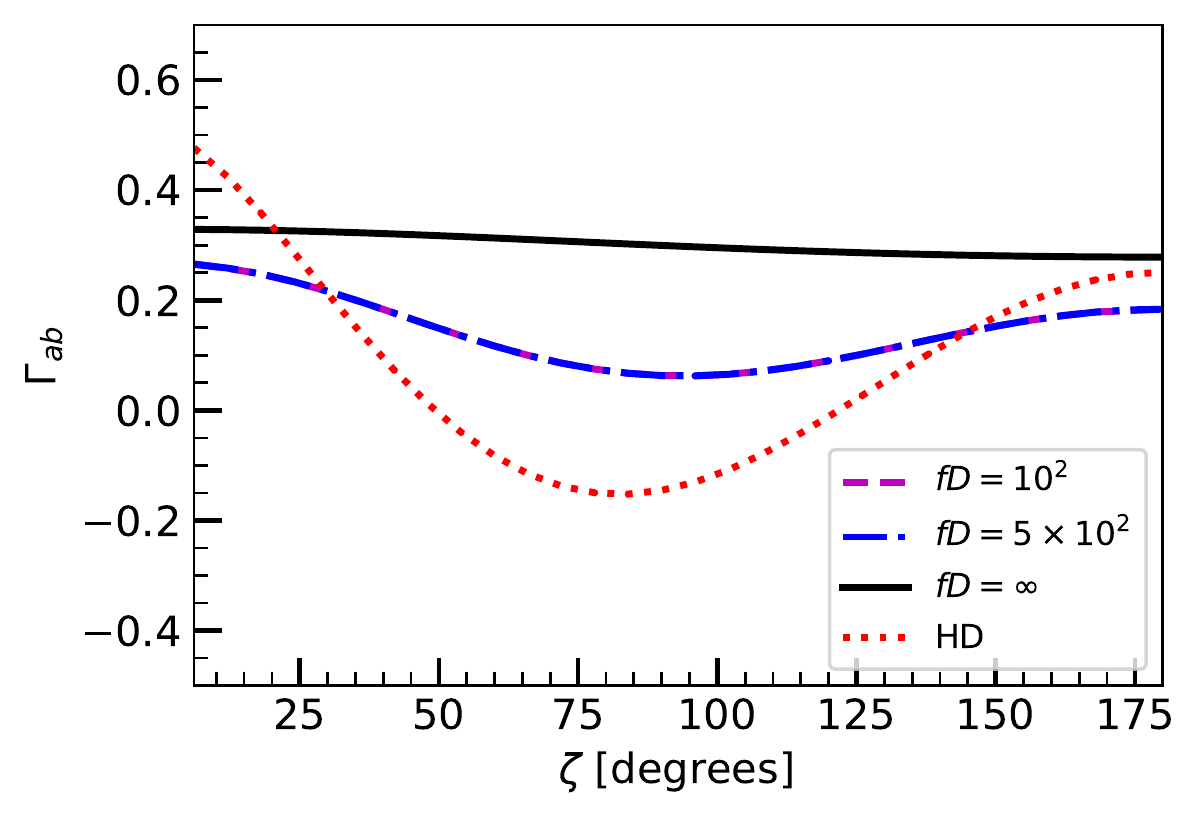}
		}
	\subfigure[ \, scalar longitudinal, $v = 1/100$ ]{
		\includegraphics[width = 0.45 \textwidth]{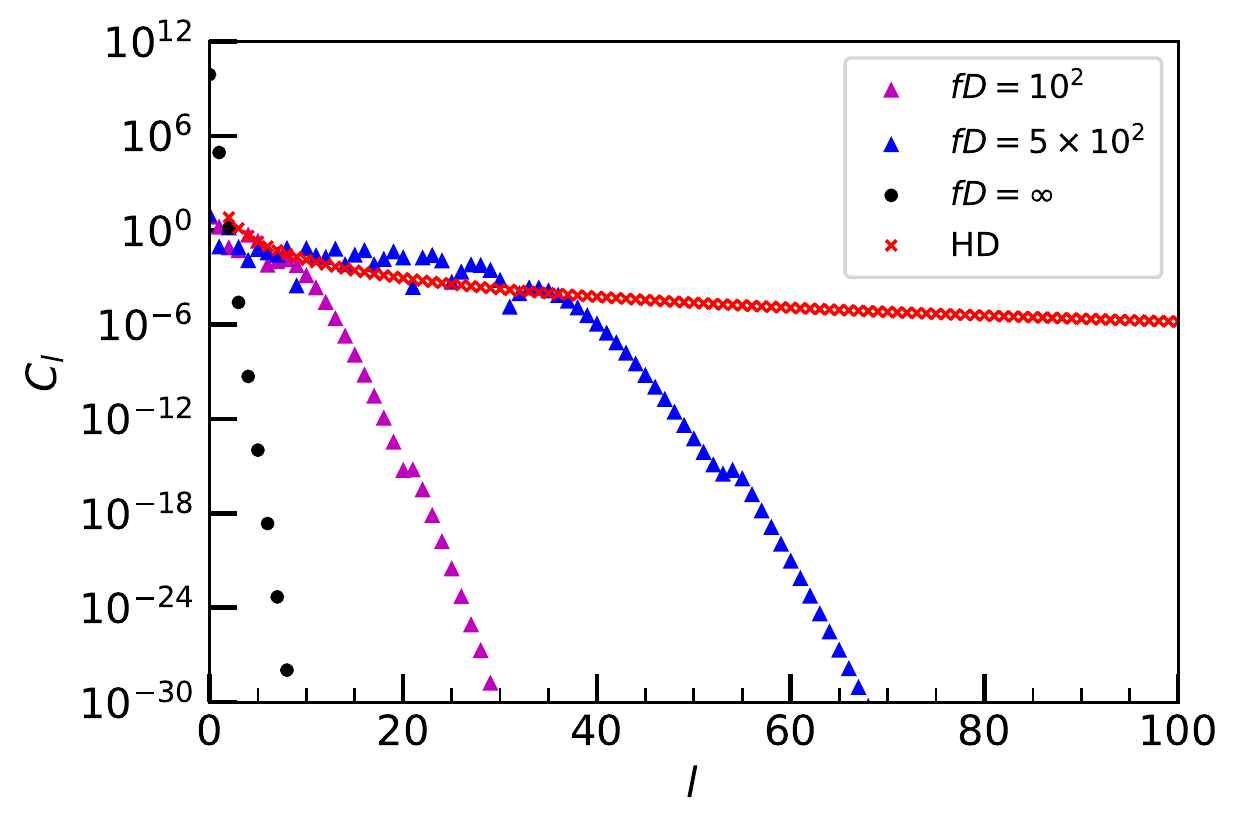}
		}
	\subfigure[ \, scalar longitudinal, $v = 1/100$ ]{
		\includegraphics[width = 0.45 \textwidth]{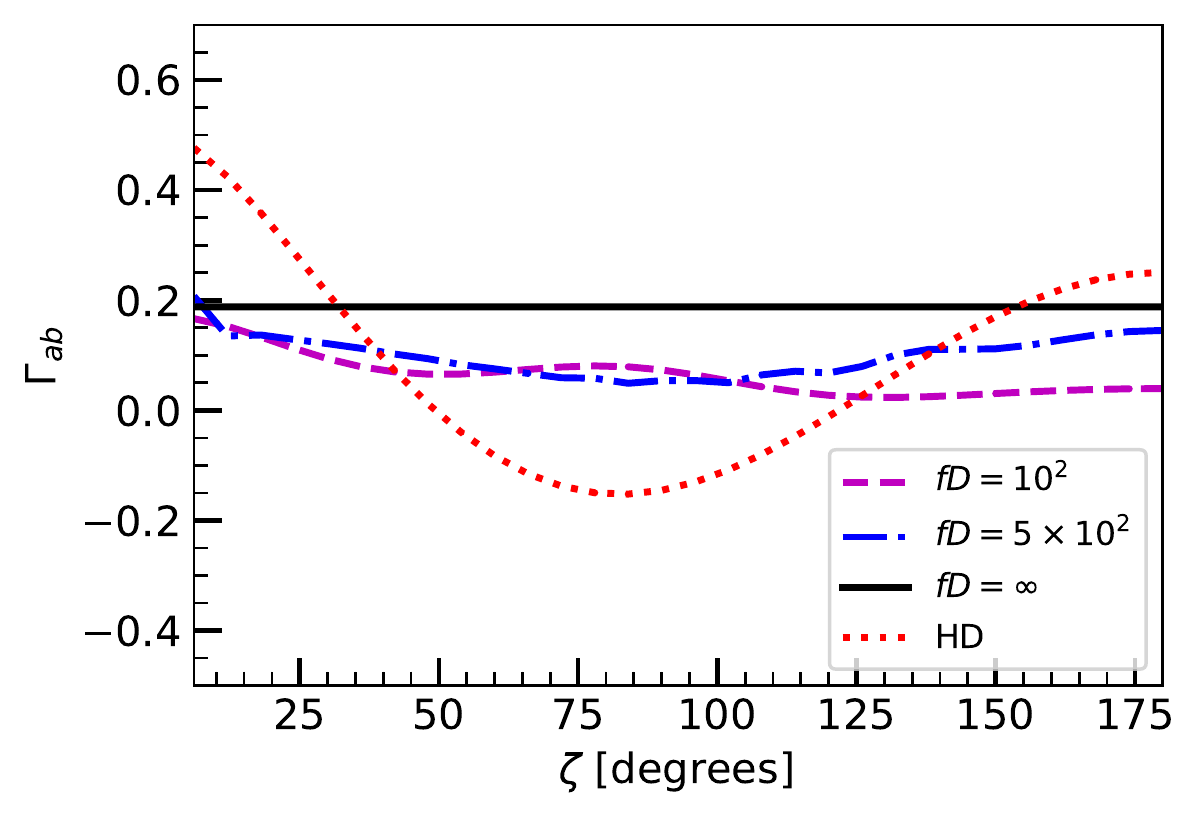}
		}
\caption{The isotropic power spectra multipoles $C_l$ and the overlap reduction function $\Gamma_{ab}(\zeta)$ of the scalar longitudinal polarization with group velocities $v = 99/100$, $v = 1/2$, and $v = 1/100$. The overlap reduction functions were constructed with only the first thirty multipoles and are normalized with respect to the Hellings-Downs (HD) correlation.}
\label{fig:ClSL}
\end{figure}

Figure \ref{fig:ClSL} shows the power spectra multipoles and the overlap reduction functions sourced by the scalar longitudinal polarization with the previous choices for the velocity, that is, near luminal, half the speed of light, and nonrelativistic. We present this together with the Hellings-Downs correlation.

The scalar longitudinal mode is undefined with $v = 1$ and infinite pulsar distances. For this reason alone, the practical advantage of keeping the pulsars at finite distances, for which it becomes possible to define the multipoles at all velocities, can be realized. Pulsars are after all real astrophysical objects birthed within our observable Universe, and so will always be at a finite distance from its constituents.

For the nearly luminal scalar (Figures \ref{fig:ClSL}(a-b)), we find the power spectrum to be dominated by the low multipoles which drop logarithmically in $l$ as $l$ increases. This leads to the particular shape of the overlap reduction function wherein most of the low multipoles contribute significantly to the angular correlation of pulsar pairs. In the infinite distance limit, however, one finds that the dipolar power is suppressed, due to which its overlap reduction function tends to be dominated by the quadrupole. The situation is strikingly different in the finite distance cases wherein the monopole and the dipole remain to be the most dominant contributions, as reflected in the overlap reduction function. We also find this monopolar and dipolar power difference between the infinite and finite pulsar distance cases to be a recurring distinguishing feature. Then, up to some point, as with the scalar transverse polarization, as one goes up the multipole ladder, the power drop is eventually disrupted, corresponding to the small scale power that is accounted for by finite pulsar distances. This is otherwise missed by taking the infinite distance limit. In the infinite pulsar distance case, due to the dipolar power drop compared to other neighboring low multipoles, the overlap reduction function between the finite distance cases and the infinite distance limit becomes distinguishable. We find that this feature is shared for other velocities.

At half the speed of light, $v = 1/2$, the scalar longitudinal multipoles feature a steeper drop as $l$ increases (Figures \ref{fig:ClSL}(c-d)). In this case, the infinite distance limit power spectrum becomes mainly dominated by the monopole and the dipole, and so the corresponding overlap reduction function at this speed shapes like the dipole. The finite pulsar distance cases instead show a significant departure from the infinite distance limit, as now their shape resembles the Hellings-Downs correlation. This can be understood by looking at the power spectrum, showing that at this speed limit, the dipolar contribution is suppressed compared to the quadrupole. Noticeably, the overlap reduction functions for the finite pulsar distance cases very much still coincide for the relevant angular separations. This changes for nonrelativistic speeds.

For the nonrelativistic scalar longitudinal polarization ($v = 1/100$), the power spectrum multipoles and overlap reduction functions are shown in Figures \ref{fig:ClSL}(e-f). In this case, in the infinite distance limit, the power spectrum is significantly dominated by the monopole, which is reflected in the overlap reduction function being  practically a horizontal line. On the other hand, for finite pulsar distances, we find most of the low multipoles continue to contribute to the power spectrum, and so produces an overlap reduction function that is visually distinguishable for the finite distance cases. We find also that at this speed limit, the dominant low multipoles determining the large angle correlations become more prominently dependent on the distance. This is reflected in both the power spectrum and the overlap reduction function. In the power spectrum, it can be seen that there are multipoles of about the same magnitude as the quadrupole, but this extends further in the multipolar index given a higher distance, thereby making the finite distance overlap reduction functions distinguishable. The same distinguishability feature occurs for the scalar transverse polarization. This opens up the possibility of distinguishing between finite distance overlap reduction functions at nonrelativistic speeds.



\end{document}